\documentclass[twocolumn,twocolappendix]{aastex63}
\usepackage{amsmath}
\usepackage{amssymb}
\usepackage{hyperref}
\hypersetup{colorlinks,breaklinks,allcolors=blue}
\usepackage[super]{nth}
\usepackage{arydshln}

\defcitealias{Genda+Abe2003}{GA03}
\defcitealias{Inamdar+Schlichting2015}{IS15}

\received{20202 February 7}
\revised{2020 May 27}
\accepted{2020 May 28}
\shorttitle{Atmospheric Erosion by Giant Impacts}
\shortauthors{Kegerreis et al.}

\begin{document}

\title{\Large Atmospheric Erosion by Giant Impacts onto Terrestrial Planets}

	\correspondingauthor{Jacob Kegerreis}
	\email{jacob.kegerreis@durham.ac.uk}
	\author[0000-0001-5383-236X]{J. A. Kegerreis}
	\affiliation{Institute for Computational Cosmology, Durham University, Durham, DH1 3LE, UK}

	\author[0000-0001-5416-8675]{V. R. Eke}
	\affiliation{Institute for Computational Cosmology, Durham University, Durham, DH1 3LE, UK}
	\author[0000-0002-6085-3780]{R. J. Massey}
	\affiliation{Institute for Computational Cosmology, Durham University, Durham, DH1 3LE, UK}
  \author[0000-0002-8346-0138]{L. F. A. Teodoro}
  \affiliation{BAERI/NASA Ames Research Center, Moffett Field, CA, USA}
  \affiliation{School of Physics and Astronomy, University of Glasgow, G12 8QQ, Scotland, UK}



\begin{abstract}

We examine the mechanisms by which atmosphere can be eroded by giant impacts onto Earth-like planets with thin atmospheres, using 3D smoothed particle hydrodynamics simulations with sufficient resolution to directly model the fate of low-mass atmospheres. We present a simple scaling law to estimate the fraction lost for any impact angle and speed in this regime. In the canonical Moon-forming impact, only around 10\% of the atmosphere would have been lost from the immediate effects of the collision. There is a gradual transition from removing almost none to almost all of the atmosphere for a grazing impact as it becomes more head-on or increases in speed, including complex, non-monotonic behaviour at low impact angles. In contrast, for head-on impacts, a slightly greater speed can suddenly remove much more atmosphere. Our results broadly agree with the application of 1D models of local atmosphere loss to the ground speeds measured directly from our simulations. However, previous analytical models of shock-wave propagation from an idealised point-mass impact significantly underestimate the ground speeds and hence the total erosion. The strong dependence on impact angle and the interplay of multiple non-linear and asymmetrical loss mechanisms highlight the need for 3D simulations in order to make realistic predictions.

\end{abstract}



\section{Introduction}
\label{sec:introduction}

Terrestrial planets are thought to form from tens of roughly Mars-sized embryos
that crash into each other after accreting from a proto-planetary disk
\citep{Chambers2001}.
At the same time, planets grow their atmospheres
by accreting gas from their surrounding nebula,
degassing impacting volatiles directly into the atmosphere, 
and outgassing volatiles from their interior \citep{Massol+2016}.

For a young atmosphere to survive
it must withstand radiation pressure of its host star,
frequent impacts of small and medium impactors,
and typically at least one late giant impact
that could remove an entire atmosphere in a single blow
\citep{Schlichting+Mukhopadhyay2018}.
In this paper, we focus on the direct, dynamical consequences 
of giant impacts onto planets like the early Earth.

Our own planet is a compelling example, 
since we can both observe an atmosphere that has survived to the present day
and be confident that a giant impact took place 
late in its evolution -- creating the Moon in the process.
Several different Moon-formation scenarios have been proposed and revised, 
but no simulations have yet resolved a crust, ocean, or atmosphere 
for the proto-Earth \citep[e.g.][]{Lock+2018,Cuk+Stewart2012}.

Focusing on the atmosphere,
the Earth's volatile abundances are remarkably different 
from those of chondrites \citep{Halliday2013}, 
which act as a record of the condensable components of the early Solar System.
Specifically, nitrogen and carbon are depleted compared with hydrogen,
which could be explained by 
the loss of N$_2$ and CO$_2$ with an eroded atmosphere
while retaining H$_2$O in an ocean \citep{Sakuraba+2019}.
Unlike the abundances, the isotope ratios match those of primordial chondrites.
Hydrodynamic escape -- 
driven by XUV radiation from the star or heat from the planet below -- 
preferentially removes lighter isotopes,
while impacts remove bulk volumes of atmosphere.
This suggests that impacts (not necessarily giant ones) 
are the primary loss mechanism,
driving fractionation by removing more atmosphere than ocean 
while preserving isotope ratios \citep{Schlichting+Mukhopadhyay2018}.

Furthermore, the relative abundances of helium and neon
in different-aged mantle reservoirs 
suggest that the Earth lost its atmosphere on at least two occasions
\citep{Tucker+Mukhopadhyay2014}.
Fractionation of xenon also indicates a complicated history of atmospheric loss
and the importance of ionic escape in addition
to impact erosion and hydrodynamic escape \citep{Zahnle+2019}.

Looking further afield, 
we have recently learnt not only that 
Earth- to Neptune-mass exoplanets are common,
but that they host a remarkable diversity of atmospheric masses
\citep{Fressin+2013,Petigura+2013,Lopez+Fortney2014}.
The stochastic nature of giant impacts
makes them a strong candidate for explaining some of the differences 
between planets that would otherwise be expected to have evolved similarly
\citep{Liu+2015b,Bonomo+2019}.
Irradiation and photoevaporation from stellar winds 
can significantly erode an atmosphere \citep{Lopez+2012,Zahnle+Catling2017}, 
but not enough to explain the diversity of planets around dim stars, 
where it should be much less effective.

Previous studies of giant-impact erosion have primarily used 
analytical approaches and 1D simulations to estimate atmospheric loss 
from a range of impact energies 
\citep[e.g.][]{Genda+Abe2003,Inamdar+Schlichting2015}.
The one-dimensional nature of these studies also means that
little work has been done on grazing collisions,
in spite of the fact that these are far more likely to occur.
Some studies have investigated oblique impacts
for much smaller (of order 10~km) objects \citep{Shuvalov2009},
in which case the erosion is only ever in the local region
and the planet's curvature is negligible.
Their results showed a strong increase in local loss for more-oblique impacts,
which is the opposite of the trend for giant impacts \citep{Kegerreis+2018}.
The typical approach for giant impacts is to estimate
the ground velocities induced by the impact
to study how much atmosphere is blown away.
This misses out the complex details of a collision
that can mix, deform, and remake both an atmosphere and the rest of the planet.
Any precise study of the consequences of a giant impact
therefore requires full 3D modelling of the
planet and atmosphere at the same time.

Recent progress has been made in the regime of thick atmospheres  
by two studies:
one with 3D simulations of head-on collisions of large super-Earths 
targeted at explaining a specific exoplanet system \citep{Liu+2015b},
and another with highly grazing impacts
that do not interact the solid layers of the planets \citep{Hwang+2018}.
This leaves serious gaps in our understanding of
the formation and atmospheric evolution of planets 
in and outside the Solar System,
both in terms of both lower atmosphere masses 
and the effect of the impact angle.

The aim of this study is thus to begin the exploration 
of this almost uncharted parameter space, 
starting in the regime of thin atmospheres.
For example: what does the impactor actually do to remove atmosphere 
in different scenarios?
How easy it is to partially erode some atmosphere as opposed to all or none?
And how do these answers change for head-on, grazing, slow, or fast impacts?

Giant impacts are most commonly studied using
smoothed particle hydrodynamics (SPH) simulations,
where planets are modelled with particles
that evolve under gravity and material pressure.
It was recently shown that at least $10^7$ (equal-mass) SPH particles 
can be required to converge on even the large-scale results
from simulations of giant impacts,
and that the resolution requirements for reliable results
depend strongly on the specific scenario and question 
\citep{Kegerreis+2019,Hosono+2017}.

Computational advances enable us for the first time to study 
the erosion of thin atmospheres with full, 3D simulations.
In this paper, we present high-resolution simulations of giant impacts
with a variety of impact angles and speeds
onto the proto-Earth,
hosting a range of low atmosphere masses.
We study the detailed mechanisms of erosion,
compare with previous analytical and 1D estimates,
and present a simple scaling law for 
the fraction of lost atmosphere in this regime.

\section{Methods} \label{sec:methods}

In this section we describe the initial conditions for the model planets,
the range of impact scenarios, 
and the previous models to which we compare our results.
The SPH simulations 
are run using the hydrodynamics and gravity code SWIFT
\footnote{
  SWIFT is in open development and publicly available at
  \href{www.swiftsim.com}{www.swiftsim.com}.
}
\citep{Kegerreis+2019,Schaller+2016}.

\subsection{Initial Conditions} \label{sec:methods:init_cond}

As a recognisable starting point, we consider an impact similar to 
a canonical Moon-forming scenario,
with a target proto-Earth of mass 0.887~$M_\oplus$
and impactor of mass 0.133~$M_\oplus$.
Both are differentiated into an iron core and rocky mantle,
constituting 30\% and 70\% of the total mass, respectively,
and have no pre-impact rotation.
The radii of the outer edge of the core and mantle
are 0.49 and 0.96~$R_\oplus$ for the target 
and 0.29 and 0.57~$R_\oplus$ for the impactor.
We use the simple \citet{Tillotson1962} iron and granite 
equations of state (EoS) \citep[][Table AII.3]{Melosh2007}
to model these materials \citep{Kegerreis+2019}
\footnote{
  Note that Appx.~B of \citet{Kegerreis+2019}
  has a typo in the sign of
  ${\rm d}u = T {\rm d}S - P {\rm d}V = T {\rm d}S + (P/\rho^2)\, {\rm d}\rho $
  just after Eqn.~B1.
}.

For the atmospheres, we use \citet{Hubbard+MacFarlane1980}'s
hydrogen--helium EoS, as described in \citet{Kegerreis+2018}.
This includes a temperature- and density-dependent specific heat capacity
and an adiabatic temperature--density relation.
An ideal gas would probably be sufficient for the smaller atmospheres,
but larger ones stray into the more-dense regime 
that this EoS is designed to include.

The Tillotson EoS does not treat phase boundaries nor mixed phases correctly
but is widely used for SPH impact simulations owing to its
computationally convenient analytical form \citep{Stewart+2019}.
These limitations could be important for studies 
that require accurate modelling of, for example, 
the thermodynamic state of low-density material in orbit.
However, for the focus in this paper on the large-scale shock-wave propagation 
and overall erosion caused by impacts, 
the details of the EoS are not expected to significantly affect the results.

The atmosphere is adiabatic above a 500~K surface, 
while the iron and silicate layers 
are given a simple temperature--density relation of $T \propto \rho^{2}$,
chosen somewhat arbitrarily to produce a central temperature of $\sim$5000~K
similar to the Earth today.
Our surface temperature is lower than the 1500~K of \citet{Genda+Abe2003},
but the fact that their erosion results are similar to those of 
\citet{Inamdar+Schlichting2015} that used many thousands of Kelvin 
suggests that the loss is not highly sensitive to this choice,
as we test directly ourselves in \S\ref{sec:results:tests}.

We test a range of atmosphere masses on the proto-Earth,
namely $10^{-1}$, $10^{-1.5}$, $10^{-2}$, and $10^{-2.5}$~$M_\oplus$,
as the lowest mass that we might expect to resolve adequately 
with $10^7$ equal-mass SPH particles.
The corresponding pressure at the base of the atmosphere is 
5.5, 2.4, 0.92, and 0.32~GPa, respectively.
They extend out to a pressure of $\sim$0.1~MPa
at 1.55, 1.27, 1.13, and 1.06~$R_\oplus$, respectively.
The Earth's atmosphere today has a mass of $\sim$$10^{-6}$~$M_\oplus$,
though it may have been much thicker in the past.

To produce the radial density and temperature profiles for each atmosphere mass, 
the surface temperature is kept fixed at 500~K for simplicity
while the surface pressure is varied until 
the desired atmospheric mass is obtained.
In other words, 
the inner two layer profiles are integrated inwards from the surface
\citep[see][Appx. A]{Kegerreis+2019},
then the atmosphere layer profile is integrated outwards,
until reaching a negligible minimum density of 10~kg~m$^{-3}$.
Separately, the total radius is also iterated
to obtain the 30:70 mass ratio of iron to rock.

Particles are then placed to precisely match these profiles using the 
stretched equal-area (SEA) method
\footnote{
  The SEAGen code is publicly available at
  \href{https://github.com/jkeger/seagen}{github.com/jkeger/seagen}
  and the python module \texttt{seagen} can be installed directly with
  \href{https://pypi.org/project/seagen/}{pip}.
}
described in \citet{Kegerreis+2019}.
This results in a relaxed arrangement of particles 
that have SPH densities within 1\% of the desired profile values,
mitigating the need for extra computation
that is otherwise required to produce initial conditions
that are settled and ready for a simulation.

\subsection{Impact Simulations} \label{sec:methods:simulations}

\begin{figure}[t]
	\centering
	\includegraphics[width=\columnwidth, trim={11mm 251mm 118mm 12mm}, clip]{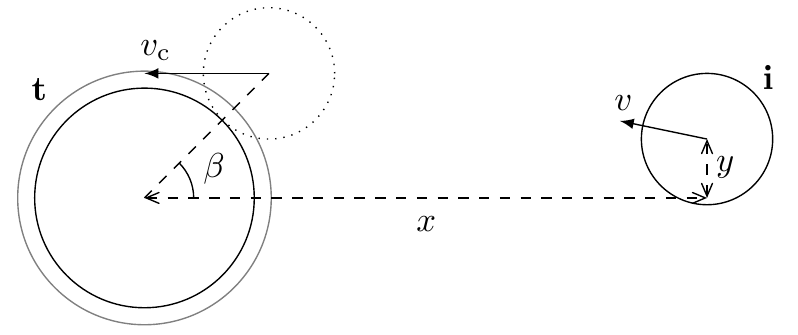}
	\caption{The initial conditions for an impact scenario, with the target
    (t) on the left and the impactor (i) on the right, 
    in the target's rest frame.
    The angle of first contact, $\beta$, is set ignoring the atmosphere
    and neglecting any tidal distortion before the collision.
    The initial separation is set by the time to impact,
    as described in Appx.~\ref{sec:init_cond}.
    \label{fig:impact_scenario}}
\end{figure}

\begin{figure}[t]
	\centering
	\includegraphics[width=\columnwidth, trim={8mm 8mm 8mm 8mm}, clip]{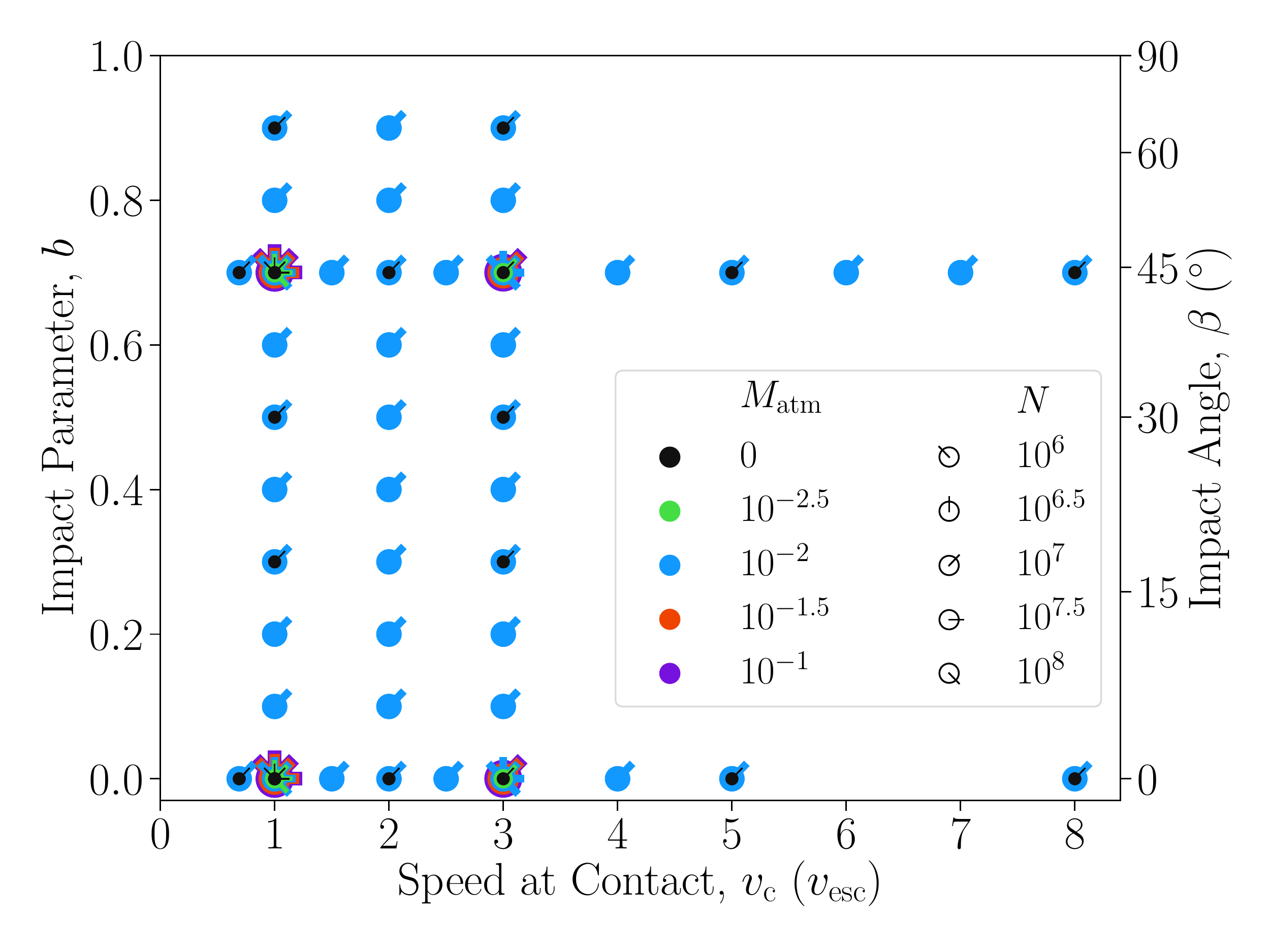}
	\caption{
    The suite of simulation scenarios,
    arranged by their speed and impact parameter at contact 
    (see Fig.~\ref{fig:impact_scenario} and Appx.~\ref{sec:init_cond}).
    As shown in the legend,
    the nested marker colours indicate the mass of the atmosphere
    (in Earth masses) 
    for each simulation,
    while the line angles indicate the number of particles per Earth mass.
		\label{fig:scenarios}}
\end{figure}

We specify each impact scenario by the impact parameter, $b=\sin(\beta)$,
and the speed, $v_{\rm c}$, at first contact of the impactor 
with the target's surface, 
as illustrated in Fig.~\ref{fig:impact_scenario}.
The initial position of the impactor is set such that 
contact occurs 1~hour after the start of the simulation, 
to allow for some natural tidal distortion 
and to not disrupt the system by suddenly introducing the large impactor 
right next to the no-longer-in-equilibrium target,
as described in Appx.~\ref{sec:init_cond}.
Note that the speed at contact is always chosen in units of 
the mutual escape speed of the system, 
$v_{\rm esc} = \sqrt{2 G \left(M_{\rm t} + M_{\rm i}\right) / 
\left(R_{\rm t} + R_{\rm i}\right)}$,
where $R_{\rm t}$ neglects the thickness of any atmosphere,
which is slightly faster for the planets with more massive atmospheres
(9.1 up to 9.6~km~s$^{-1}$).

We run a primary suite of 
74 simulations with $\sim$$10^7$ SPH particles,
plus 10 of these scenarios re-simulated additionally with 
$10^6$, $10^{6.5}$, $10^{7.5}$, and $10^8$ particles
for convergence tests,
plus 12 miscellaneous tests with $10^7$ particles 
detailed in \S\ref{sec:results:tests}.
To be precise, these stated particle numbers refer to the 
number of particles per Earth mass 
(the bare target plus impactor mass is 1.02~$M_\oplus$).
Thus, the numerical resolution stays the same for simulations 
with different-mass planets.
For example, a `$10^7$' simulation
that includes a 0.1~$M_\oplus$ atmosphere actually contains a total 
of $\sim$1.12$\times 10^7$ particles.
For most of the suite we focus on the $10^{-2}$~$M_\oplus$ atmosphere.

Fig.~\ref{fig:scenarios} summarises the parameters for each simulation.
Note that the ${v_{\rm c} = 0.75}$~$v_{\rm esc}$ scenarios 
would require some third body to have slowed down the impactor 
during its approach to below the mutual escape speed.
This is unlikely in the case of 
primary impactors falling in to the Earth in our solar system,
but is a useful test for the consequences of a highly grazing impact 
resulting in a large bound fragment that will re-impact at a later time.
It also lets us compare with other models,
which predict little erosion in this regime.

At the high-speed end, given the Earth's position in the Solar System,
5~$v_{\rm esc}$ is around the highest typical velocity that might be expected 
for an impact \citep{Raymond+2009}.
For context, the Earth's orbital speed around the Sun is about 3~$v_{\rm esc}$.
The suite's extension to 8~$v_{\rm esc}$ both allows us to test 
the extreme end of the parameter space 
and is a regime that could be more common in other planetary systems,
for example with a more massive star or a 
target planet deeper in the star's potential well.
Furthermore, in studies like that by \citet{Inamdar+Schlichting2015}
where erosion is estimated as a function of the impactor's momentum,
using very high velocities will allow us to test 
the degeneracy between impactor mass and speed 
across a wide range of momenta
in future suites with different impactor masses.
For the relatively small impactor mass used here,
even 8~$v_{\rm esc}$ is not predicted by \citet{Inamdar+Schlichting2015}
to remove more than 3/4 of the atmosphere.

The simulations are run using SWIFT
with a simple `vanilla' form of SPH 
plus the \citet{Balsara1995} switch for the artificial viscosity
as described in \citet{Kegerreis+2019}
to a simulation time of 100,000~s (roughly 28~hours)
in a cubic box of side 80~$R_\oplus$ to allow the tracking of ejecta.
Any particles that leave the box are removed from the simulation.
Throughout the first 10~hours we record snapshots every 100~s,
for high time resolution during the impact and its immediate aftermath.
To reduce data storage requirements,
we then output snapshots every 1000~s for the remainder.

\begin{figure*}[t]
	\centering
	\includegraphics[width=0.94\textwidth, trim={54mm 29.5mm 77mm 9.5mm}, clip]{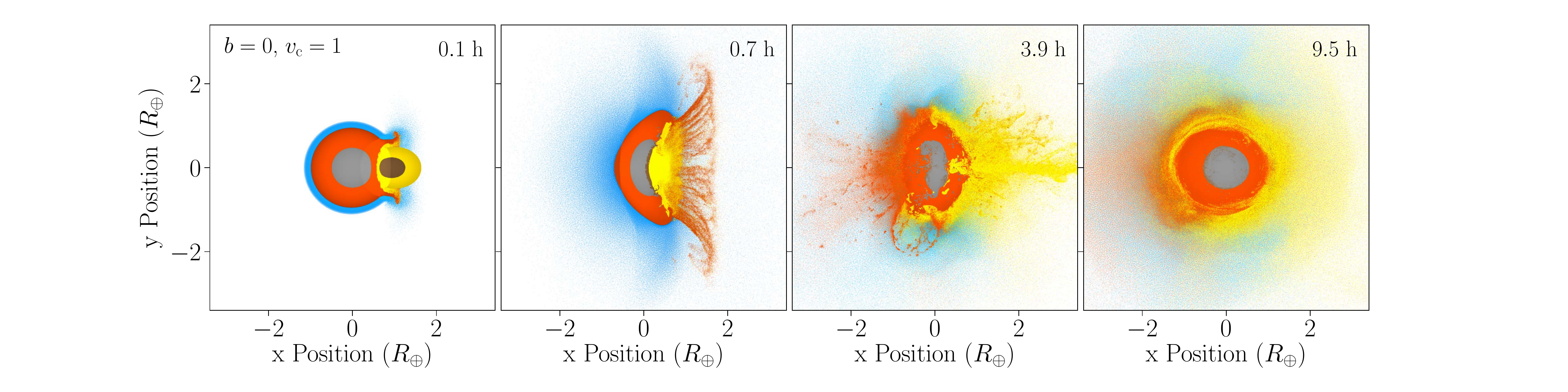}\\\vspace{-0.4mm}
	\includegraphics[width=0.94\textwidth, trim={54mm 29.5mm 77mm 9.5mm}, clip]{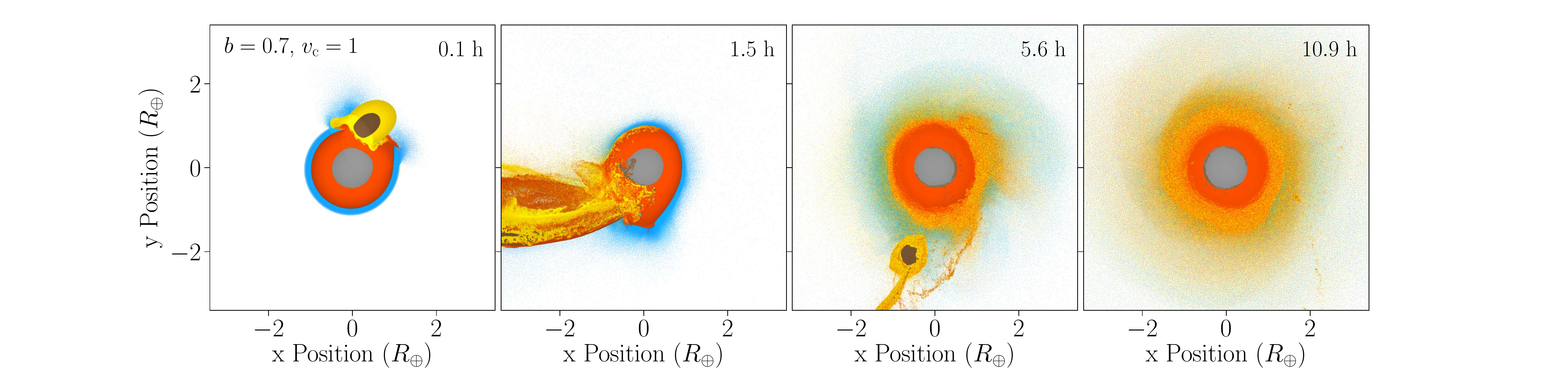}\\\vspace{-0.4mm}
	\includegraphics[width=0.94\textwidth, trim={54mm 29.5mm 77mm 9.5mm}, clip]{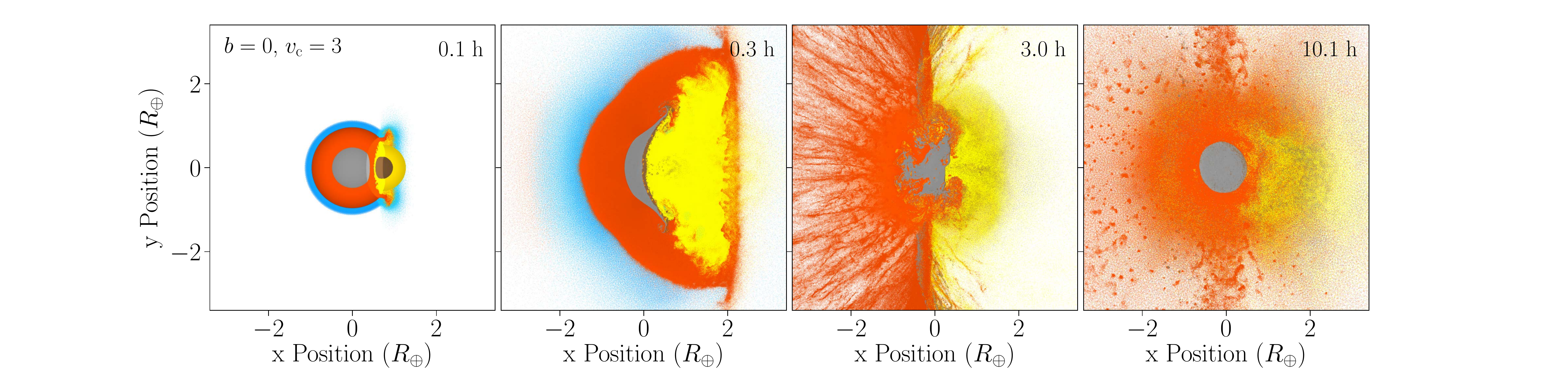}\\\vspace{-0.4mm}
	\includegraphics[width=0.94\textwidth, trim={54mm 8mm 77mm 9.5mm}, clip]{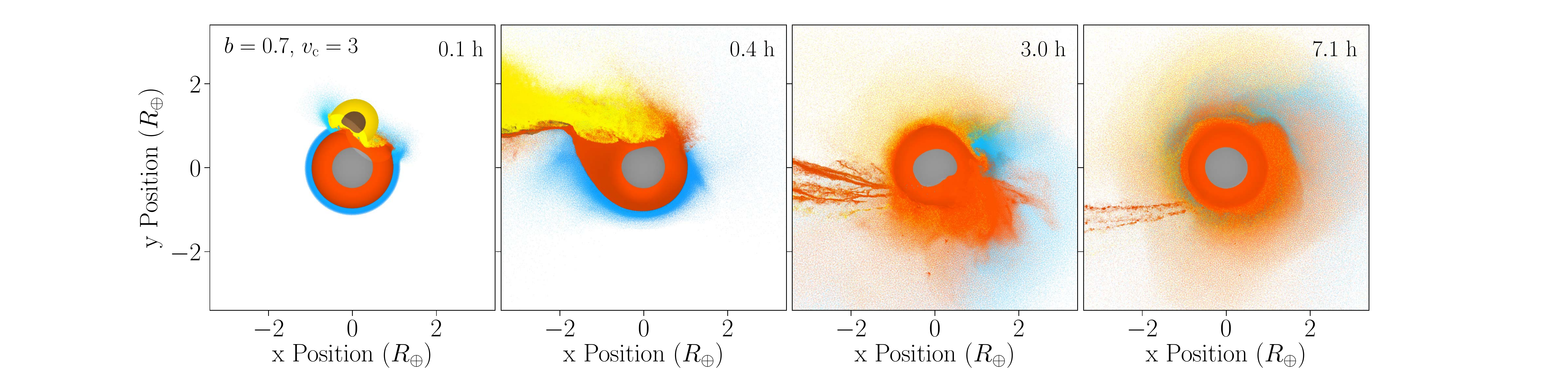}\\
  \caption{
    Illustrative early snapshot cross-sections 
    from the four fiducial impact simulations -- 
    head-on and slow, grazing and slow, head-on and fast, grazing and fast --  
    with ${b=0}$ or $0.7$, and $v_{\rm c}=1$ or $3$ 
    (labelled throughout in units of $v_{\rm esc}$),
    with the 1\%~$M_\oplus$ atmosphere and $\sim$$10^8$ SPH particles.
    Grey and orange show the target's core and mantle material respectively,
    and brown and yellow show the same for the impactor.
    Blue is the target's atmosphere.
    The colour luminosity varies slightly with the internal energy.
    Note that the snapshots are at different times for each simulation 
    to show the evolution in each case.
    The impactors are travelling in the $-$$x$ direction 
    at the moment they contact the target 
    (see Fig.~\ref{fig:impact_scenario}).
    Animations of the early evolution of these impacts are available at
    \href{http://icc.dur.ac.uk/giant_impacts/atmos_fid_1e8_anim.mp4}{icc.dur.ac.uk/giant\_impacts}.
		\label{fig:fiducial_snaps}}
\end{figure*}

\begin{figure*}[t]
	\centering
	\includegraphics[width=0.95\textwidth, trim={54mm 8mm 77mm 9.5mm}, clip]{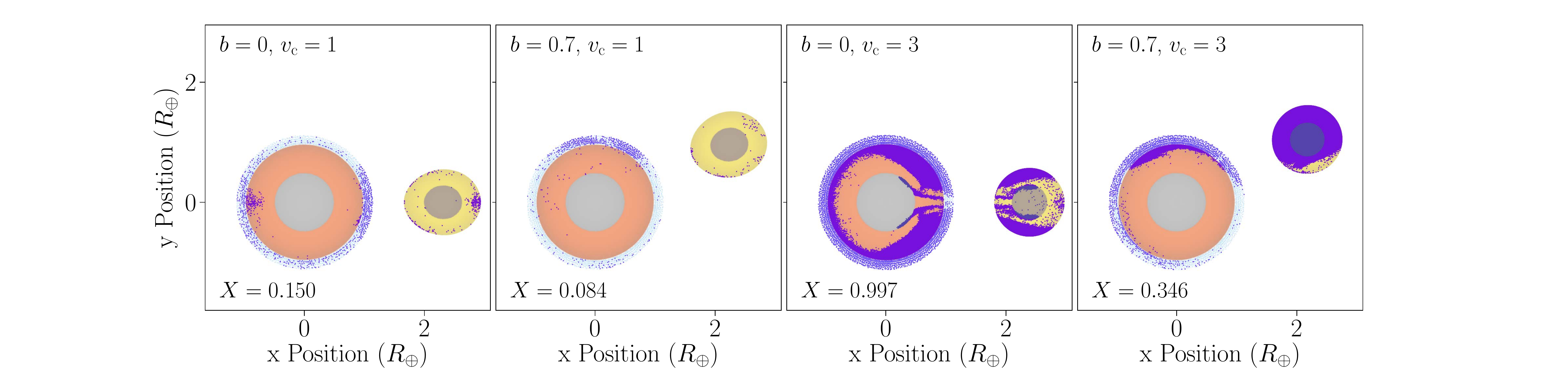}
  \caption{
    The particles that will become unbound and escape the system,
    highlighted in purple on a pre-impact snapshot,
    for the four fiducial impacts
    and our standard $\sim$$10^7$ SPH particles.
    The other particle colours are muted versions of those in 
    Fig.~\ref{fig:fiducial_snaps}
    as a background for the highlighted ones.
    Only a thin cross-section of the particles that are 
    within one SPH smoothing length of $z=0$ are shown for clarity.
    For grazing impacts, higher latitudes may suffer less erosion
    (see \S\ref{sec:results:ground}).
    The $X$ values give the total mass fraction of the atmosphere that is lost.
		\label{fig:fid_will_be_unb}}
\end{figure*}

\subsection{Analytical and 1D Models} \label{sec:intro:models}

We use two previous erosion studies for comparison with our 3D simulations,
both for the resulting loss of atmosphere 
and for the shock waves caused by the impact.
\citet{Genda+Abe2003} used 1D models
to simulate the reaction of the atmosphere 
to a shock from vertical ground motion.
Their results for the local fraction of lost atmosphere, $X_{\rm local}$, 
are fitted well by a simple linear function of the ground speed, $v_{\rm gnd}$,
in units of the escape velocity:
${X_{\rm local} = - 1/3 + 4/3 \; (v_{\rm gnd}/v_{\rm esc})}$
capped at zero and one (their Eqn.~17),
which they conclude is largely insensitive 
to the initial conditions of the atmosphere.

\citet{Inamdar+Schlichting2015} performed similar 1D, Lagrangian,
vertical-shock simulations, 
but extended them to include thicker atmospheres
up to 10\% of the solid mass of the planet.
They agree with \citet{Genda+Abe2003} for thin atmospheres.
\citet{Schlichting+2015} also created a model for predicting 
the ground speeds caused by a giant impact,
by treating the collision as a point-mass explosion
on a spherical planet of constant density.
They assumed momentum conservation
with a uniform speed in the spherical region traversed by the shock front, 
which leads to the vertical ground speed as a function of distance, $l$,
from the impact point:
$v_{\rm gnd} = v_{\rm imp}(M_{\rm i}/M_{\rm t}) [(l/(2R_{\rm t}))^2(4 - 3l/(2R_{\rm t}))]^{-1}$ 
(their Eqn.~28),
where $v_{\rm imp}$ is the speed of the impactor
and $R_{\rm t}$ and $M_{\rm t}$ are the radius and mass of the target planet.
By combining their speed estimates with the 1D local erosion model,
they presented predictions for the global atmospheric mass-loss fraction
as a function of the impactor speed and velocity
for different atmosphere masses (their Fig.~5).

\begin{figure*}[t]
	\centering
	\includegraphics[width=\textwidth, trim={85mm 33.5mm 108mm 9.5mm}, clip]{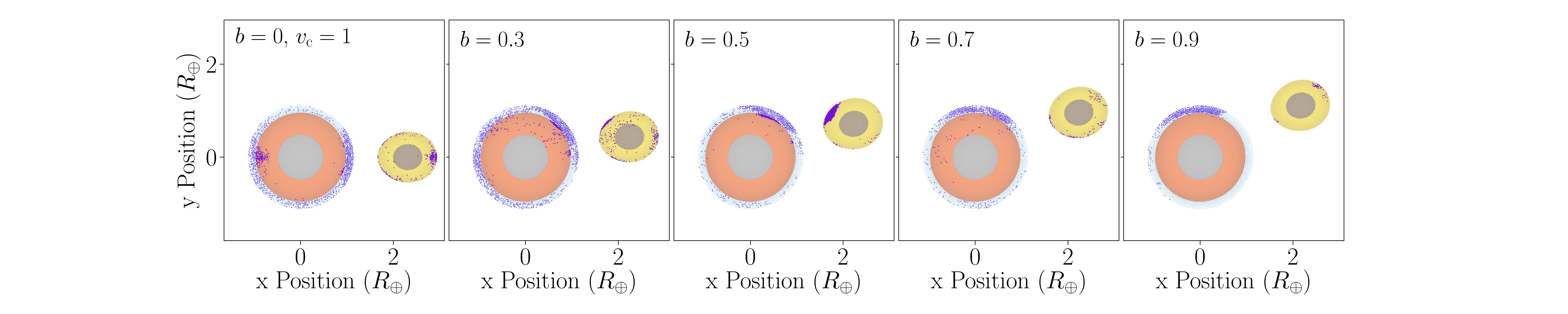}\\\vspace{-0.4mm}
	\includegraphics[width=\textwidth, trim={85mm 33.5mm 108mm 9.5mm}, clip]{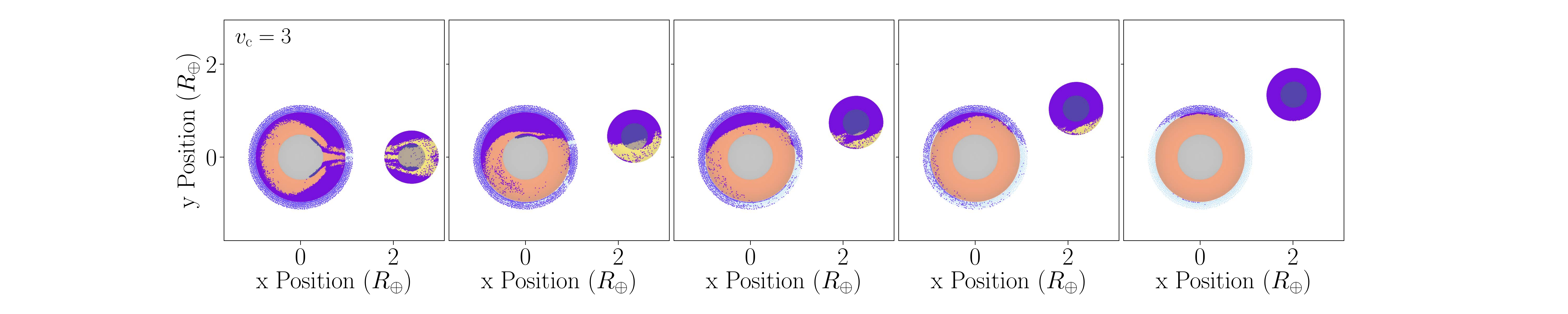}\\\vspace{-0.4mm}
	\includegraphics[width=\textwidth, trim={85mm 33.5mm 108mm 9.5mm}, clip]{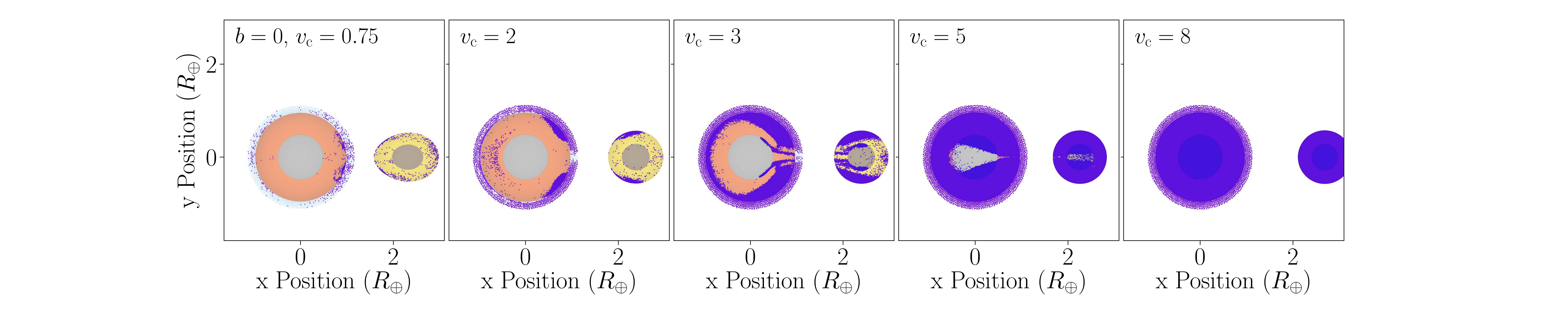}\\\vspace{-0.4mm}
	\includegraphics[width=\textwidth, trim={85mm 8mm 108mm 9.5mm}, clip]{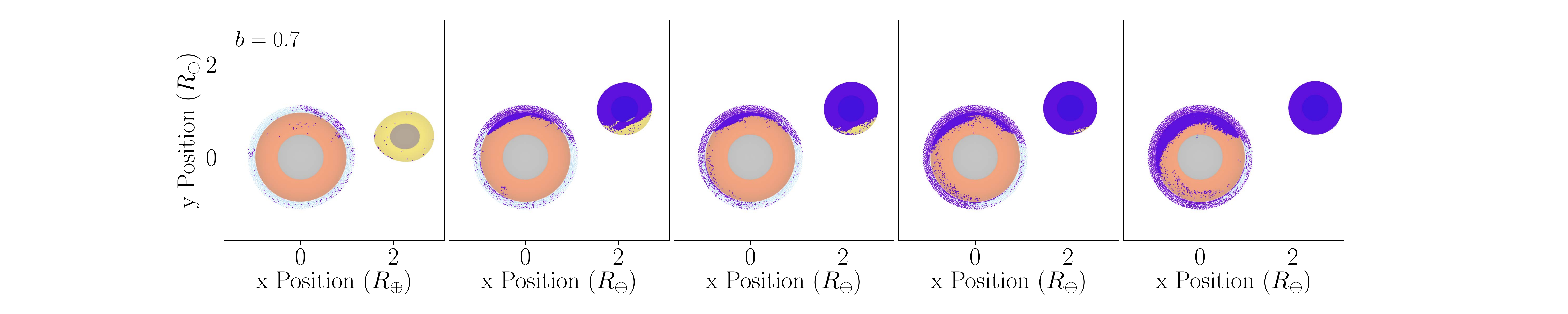}
  \caption{
    The particles that will become unbound and escape the system,
    as in Fig.~\ref{fig:fid_will_be_unb},
    for example subsets of (top two rows) different impact parameters 
    and (bottom two rows) different speeds at contact.
		\label{fig:will_be_unb_b_v}}
\end{figure*}

\section{Results and Discussion} \label{sec:results}

We begin investigating the simulations
with an overview of the general features and consequences 
of these classes of impacts.
Then, we focus on the isolated effects of changing 
the impact parameter, speed, or atmosphere mass,
and examine the time at which material is ejected.
We consider the ground speeds and localised loss
to compare our results with previous estimates,
then collate all the simulation results to find a simple scaling law
for the total atmospheric erosion from any scenario in this regime.

\subsection{General Features of Impacts and Erosion} \label{sec:results:overview}

We choose four simulations to act as fiducial comparisons 
for the rest of the suite, 
demonstrating head-on and grazing, slow and fast scenarios.
They stand out in Fig.~\ref{fig:scenarios} as the impacts 
for which we simulate multiple atmosphere masses and with multiple resolutions.
Snapshots from these fiducial simulations
are shown in Fig.~\ref{fig:fiducial_snaps},
for a target with a 1\%~$M_\oplus$ atmosphere, 
using $\sim$$10^8$ SPH particles.

In general, the impactor merges with the target for head-on or slow cases,
but may not for fast, grazing impacts.
In addition to any differences in the resulting fraction of lost atmosphere, 
the timing and cause of loss can also vary significantly 
with the impact scenario.
For example, atmosphere may be eroded by -- 
in approximately chronological order:
\begin{itemize} 
  \item Direct encounter with the very-much-not-a-point-mass 
  impactor passing through,
  most dramatically demonstrated in the high-speed, grazing case (\nth{4} row);
  \item The shock wave travelling through the planet from the impact point,
  which even erodes some mantle as well 
  in the high-speed, head-on case (\nth{3} row);
  \item Subsequent oscillations of the planet,
  such as the plume of impactor mantle in the \nth{3} snapshot 
  of the low-speed, head-on case (\nth{1} row) --
  much like the large splash created after dropping a stone into a pond;
  \item The secondary impact of the impactor 
  following an initial grazing collision,
  as in the \nth{3} snapshot of the low-speed, grazing case (\nth{2} row).
\end{itemize} 
All of these mechanisms may contribute to the total loss in a given scenario.
This provides some context with which to consider the rest of the suite
and some appreciation for the complexity created 
by all these processes intermingling. 

The particles that are eroded by these four impacts
are highlighted in Fig.~\ref{fig:fid_will_be_unb},
selected by being gravitationally unbound and remaining so 
until the end of the $10^5$~s simulation or 
until the time the particle exits the 80~$R_\oplus$-wide simulation box.
The resulting mass fractions of lost atmosphere are 
0.15, 0.08, 1.0, and 0.39, respectively.
We revisit these final loss results in the context of the whole suite
after presenting the rest of the simulations
and introducing the previous analytical and 1D estimates for comparison.
For now, Fig.~\ref{fig:fid_will_be_unb} demonstrates
the expected qualitative results 
following the above discussion of Fig.~\ref{fig:fiducial_snaps}:
the head-on, slow case loses atmosphere 
around the impact point and the antipode;
the grazing, slow case shows little antipode erosion, suggesting a weaker shock,
and primarily loses atmosphere in the direct path of the impactor;
the head-on, fast impactor has blasted off 
almost all the atmosphere and some mantle from the strong shock wave;
and the grazing, fast case is similar to the grazing, slow one, 
but the impactor has taken some of the mantle 
in its path along with the atmosphere
and blasted away some atmosphere around the antipode.
The grazing, fast impactor itself also remains unbound 
in this hit-and-run collision.

\begin{figure*}[t]
	\centering
	\includegraphics[width=0.80\textwidth, trim={7mm 8mm 9mm 8mm}, clip]{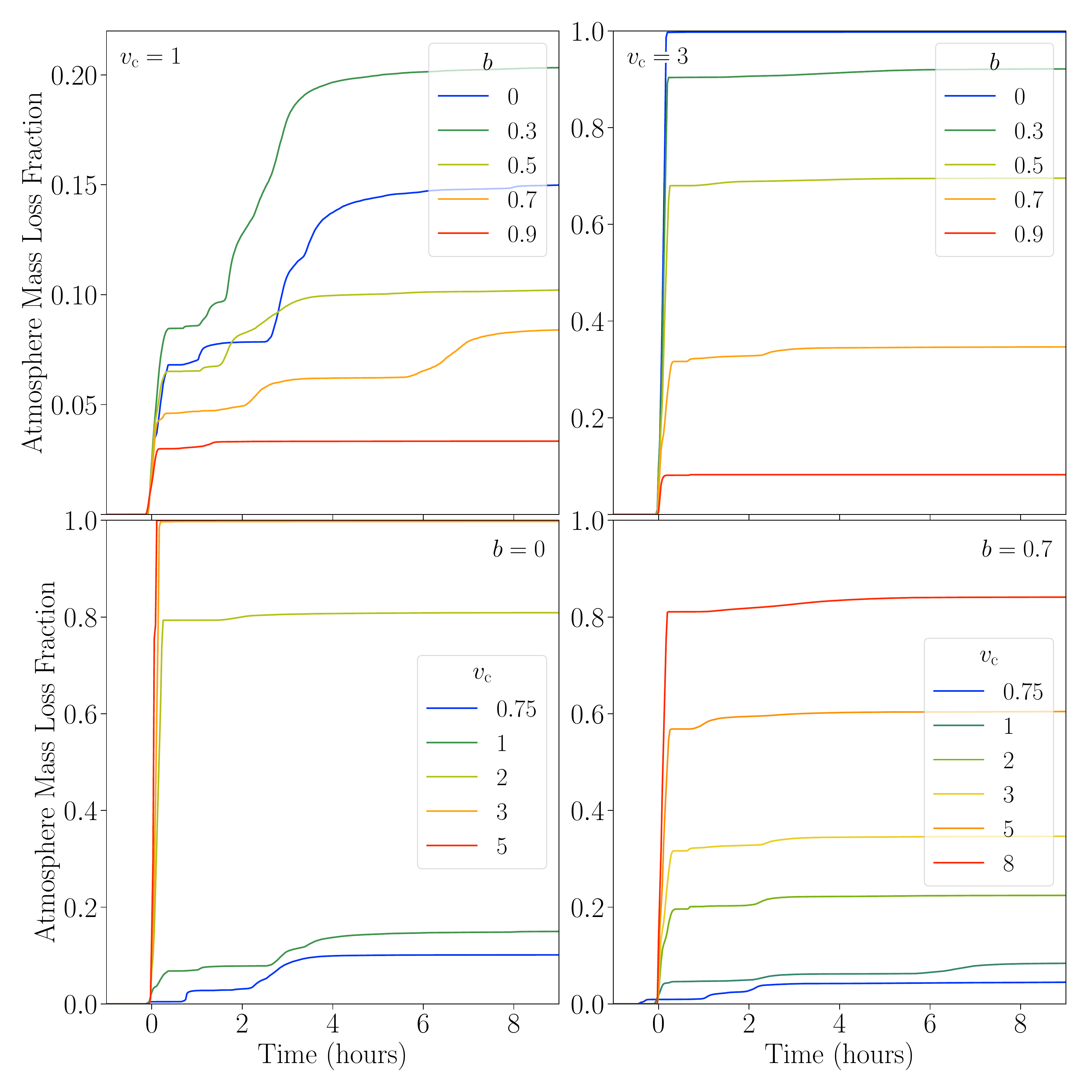}
  \caption{
    The early time evolution of the mass fraction of unbound atmosphere
    for different subsets of impact parameters and speeds 
    (labelled in units of $v_{\rm esc}$)
    with the 1\%~$M_\oplus$ atmosphere and $\sim$$10^7$ particles.
    i.e. the times at which the highlighted atmosphere particles in 
    Fig.~\ref{fig:will_be_unb_b_v} become unbound.
    Note that the vertical axis in the top-left panel does not reach 1.
    Time $=0$ is set to be the time of contact
    from Appx.~\ref{sec:init_cond},
    1~hour after the start of the simulation.
		\label{fig:m_unb_evol_b__v__}}
\end{figure*}

\begin{figure*}[t]
	\centering
	\includegraphics[width=0.8\textwidth, trim={7mm 8mm 9mm 8mm}, clip]{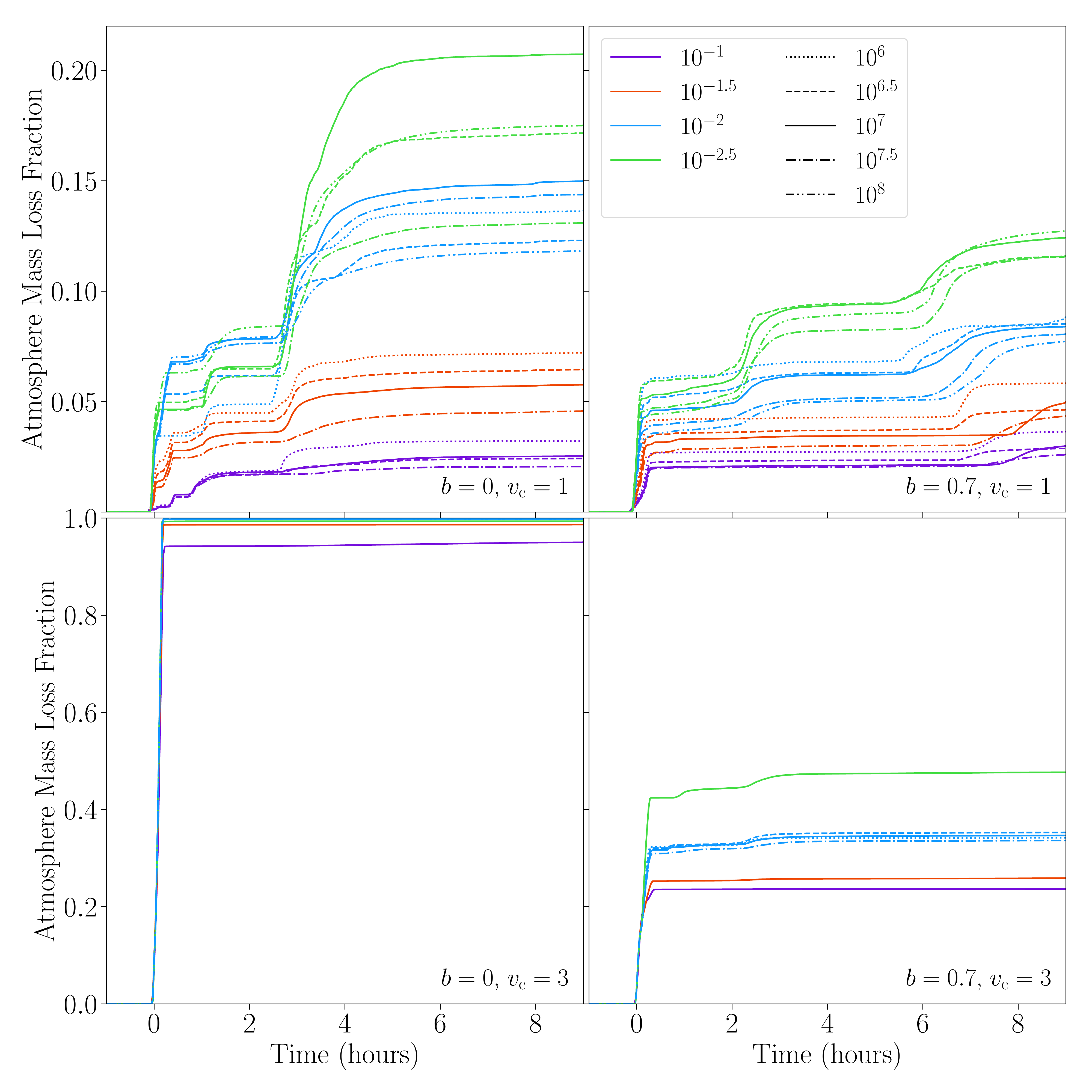}
  \caption{
    The early time evolution of the mass fraction of unbound atmosphere
    for the four fiducial impact scenarios
    with different atmosphere masses 
    (labelled in units of $M_\oplus$).
    Note that the vertical axes in the top panels do not reach 1.
    The dotted and dashed lines show the loss evolution 
    for different numbers of particles, 
    as given in the legend.
		\label{fig:m_unb_evol_m_atm}}
\end{figure*}

Note that even head-on collisions are not 
\emph{perfectly} rotationally symmetric in our simulations,
because the system is represented using a finite number of particles.
For example, in addition to the large plume of material ejected during 
the low-speed, head-on impact, 
a small blast occurs on the $-$$y$ side 
(\nth{1} row in Fig.~\ref{fig:fiducial_snaps}).
This is impactor material that initially plunges deep into the target's centre.
Being much less dense than the iron core, it gets rapidly forced back out
in a random direction determined by the arrangement of the discrete particles.
In our simulations of the same impact scenario
using different numbers of particles,
the same eruption of material is produced at the same time,
but with different random orientations in the $y$--$z$ plane.
On the one hand, this highlights the imperfect symmetry of our SPH planets,
which prevents the modelling of perfectly idealised head-on collisions.
On the other hand, this also demonstrates 
the importance of using fully 3D hydrodynamical simulations 
to study realistically chaotic giant impacts,
where we should expect some level of asymmetry
and precisely head-on impacts have a probability of zero.
At any rate, this feature ejects negligible unbound material,
so does not affect the overall results of this specific study.

We now turn to the rest of the suite in a similar manner,
continuing this initial overview of general behaviour.
The top two rows of Fig.~\ref{fig:will_be_unb_b_v} 
highlight the particles that become lost 
from subsets of changing-impact-parameter scenarios,
with either the low or high fiducial speeds
and the same atmosphere and number of particles.
Filling in the gaps between the fiducial examples, 
there is a trend from more global, shock-driven erosion 
for low impact parameters, 
to direct, localised erosion for high impact parameters.

Fig.~\ref{fig:will_be_unb_b_v}'s bottom two rows show the eroded particles 
from subsets of changing-speed scenarios,
with either the head-on or grazing fiducial impact parameters.
Even though the slowest impactors make contact at below the escape speed,
they still erode some atmosphere locally.
For head-on impacts, by $v_{\rm c} = 2$~$v_{\rm esc}$, 
already almost all of the atmosphere is eroded.
At higher speeds, more mantle is also lost, 
and $v_{\rm c} = 8$~$v_{\rm esc}$ disintegrates the planet entirely.
The faster grazing impacts can still deliver enough energy 
to drive some antipodal loss but
remove systematically less atmosphere than head-on collisions,
and even by $v_{\rm c} = 5$~$v_{\rm esc}$
with $b=0.7$ almost half of the atmosphere still survives.

We find broadly similar behaviour for different masses of atmosphere,
in simulations with the same fiducial impact parameters and speeds.
For slow, head-on impacts onto targets with 
atmospheres at and below $\sim$$10^{-2}$~$M_\oplus$,
the mantle erosion is similar to the case with zero atmosphere.
Thicker atmospheres begin to significantly cushion the mantle from erosion.
The low- and zero-mass atmosphere cases are also similar in the 
other three fiducial scenarios, 
although for slow, grazing collisions the 
thicker atmospheres can affect the path of 
the impactor as it passes through, making the comparison less direct.
At higher speeds, the atmosphere mass makes less difference,
especially in the head-on case,
because both any gravitational acceleration 
and hydrodynamical deceleration will have smaller effects.

\begin{figure*}[t]
	\centering
	\includegraphics[width=\textwidth, trim={54mm 8mm 77mm 9mm}, clip]{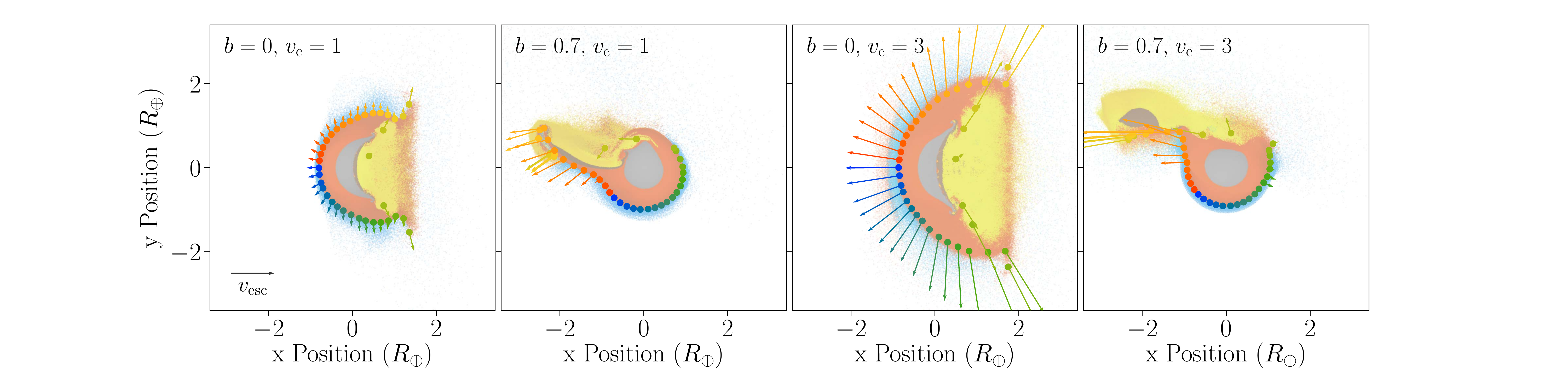}
  \caption{
    Example positions and velocities of the outermost `ground' particles 
    of the target's mantle,
    for the four fiducial simulations at 0.4, 0.6, 0.2, and 0.2~h after contact, 
    respectively.
    The colours show the particles' original longitudes in bins of $10^\circ$, 
    from $0^\circ$ (pale green) at the point of contact,
    to $180^\circ$ (red) and $-180^\circ$ (blue) at the antipode,
    here within $\pm$$5^\circ$ latitude of the $0^\circ$ impact plane.
    The maximum speeds in Fig.~\ref{fig:gnd_vel_peak}
    are taken from across all snapshot times,
    whereas only single snapshots are shown here.
		\label{fig:gnd_quiver}}
\end{figure*}

\subsection{Erosion Time Evolution} \label{sec:results:evol}

The time at which the lost atmosphere becomes unbound 
is shown in Fig.~\ref{fig:m_unb_evol_b__v__},
for subsets of changing-impact-parameter and changing-speed scenarios.
Significant atmosphere can be eroded after the initial impact,
especially for slower collisions with low impact parameters.
This corresponds to the potentially violent oscillations of the planet,
shocking away surviving shells of atmosphere 
or even ejecting plumes of material, 
as seen in the slow, head-on fiducial example
(Fig.~\ref{fig:fiducial_snaps}).
For high impact parameters, delayed erosion can also be caused by 
the secondary collision of grazing impactor fragments.
However, given the low speeds required for a grazing fragment to return
and the likely reduced mass of the fragment,
this has a smaller effect.

The majority of loss has finished 
by 4--8~hours after contact in all cases,
and the eroded mass remains constant to within a few percent 
up to the end of the 28~hour simulations.
For impact speeds of $\gtrsim 2$~$v_{\rm esc}$,
the erosion is completed almost immediately,
with little change after only the first couple of hours.
For low impact parameters, this is simply because the entire atmosphere 
is blown away by the initial shock.
For grazing collisions, it is the lack of re-impacting fragments 
that reduces any later erosion.

\begin{figure*}[t]
	\centering
	\includegraphics[width=\textwidth, trim={13mm 197mm 13mm 12mm}, clip]{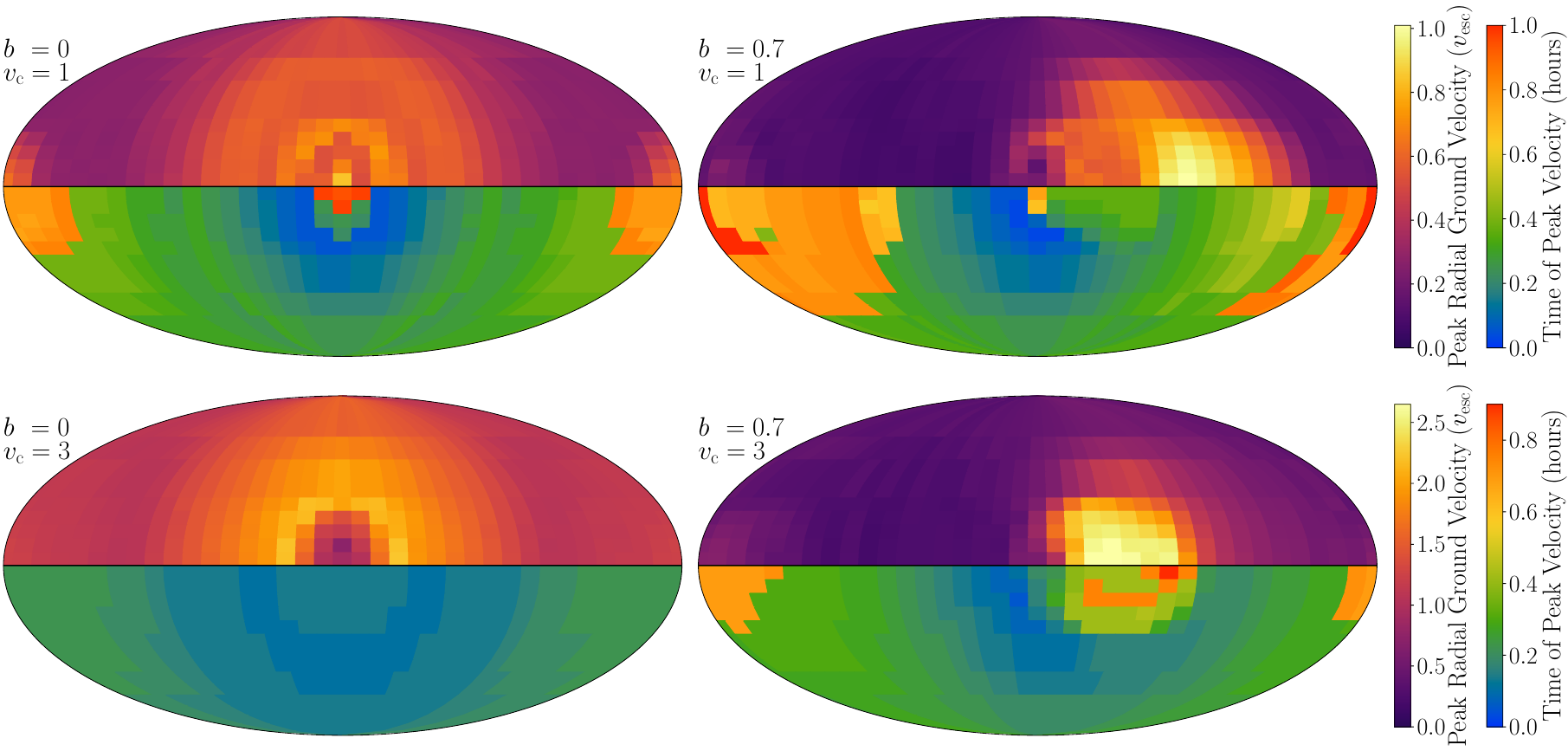}
  \caption{
    The maximum outwards radial velocity of the outermost particles
    of the target's mantle (top hemispheres)
    and the times at which they occur (bottom hemispheres)
    for the four fiducial simulations,
    on a Mollweide projection with the point of contact at $(0^\circ, 0^\circ)$
    in the centre, as described in Fig.~\ref{fig:gnd_quiver}.
    The impacts are symmetric in latitude 
    so only one hemisphere is shown for each parameter.
    Note that the top, low-speed pair simulations share the same colour bars
    that have different limits to those shared by the bottom, high-speed pair.
		\label{fig:gnd_vel_peak_time}}
\end{figure*}

\begin{figure*}[t]
	\centering
	\includegraphics[width=0.8\textwidth, trim={7mm 8mm 6mm 8mm}, clip]{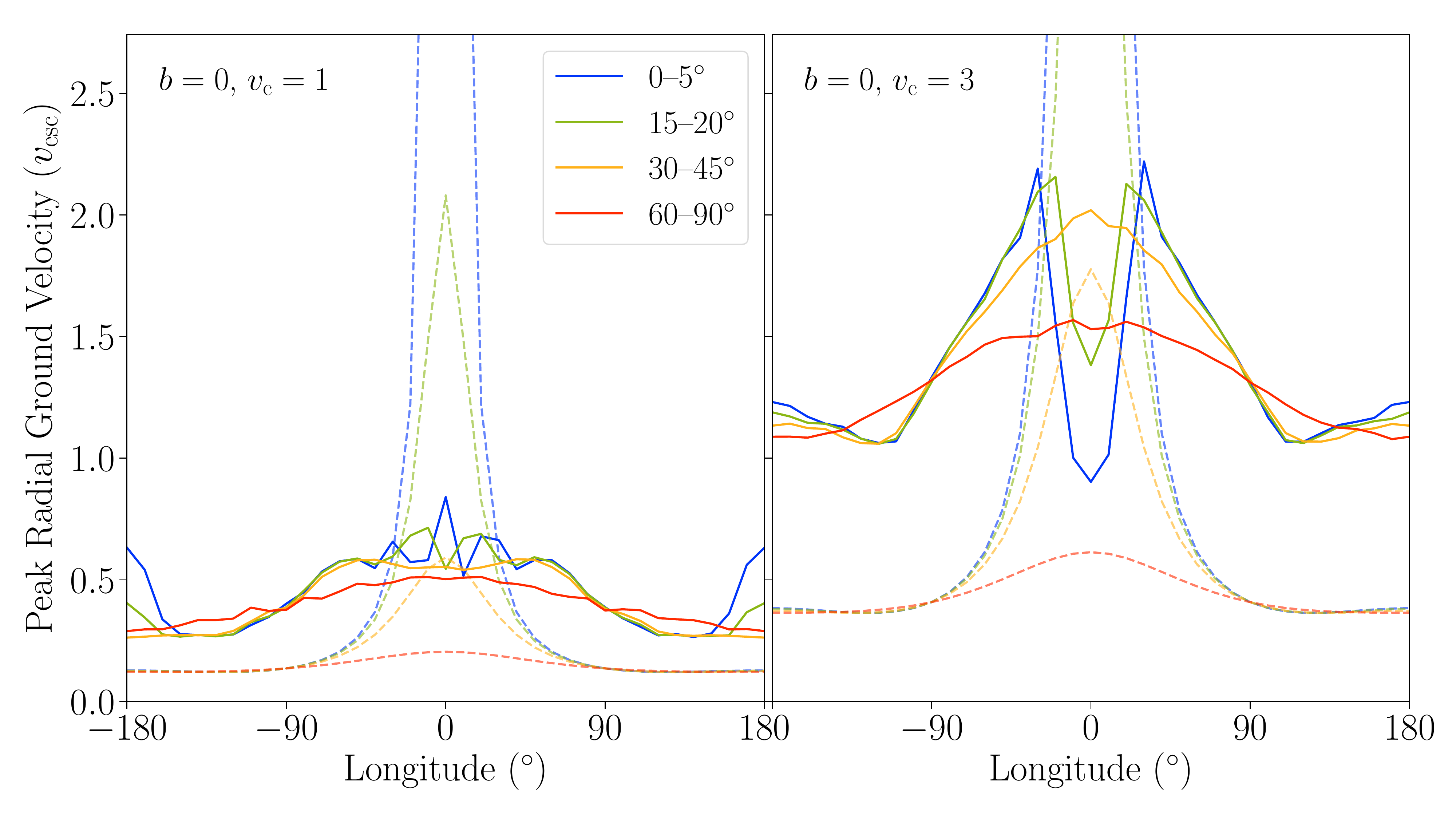}
  \caption{
    The maximum outwards radial velocity of the outermost particles
    of the target's mantle 
    as a function of longitude away from the impact point, 
    in separate, similar-area $|$latitude$|$ bins 
    -- effectively showing horizontal slices across 
    Fig.~\ref{fig:gnd_vel_peak_time} --
    for the two head-on fiducial simulations.
    The dashed lines show the estimated ground speeds at the same latitudes
    from \citet{Inamdar+Schlichting2015}, 
    based on conserving a point-mass impactor's momentum 
    in a spherical shock wave.
		\label{fig:gnd_vel_peak}}
\end{figure*}

Fig.~\ref{fig:m_unb_evol_m_atm} shows the time evolution 
for the loss of the different-mass atmospheres.
The qualitative evolution is similar in most cases,
especially for the $10^{-2}$ and $10^{-2.5}$~$M_\oplus$ atmospheres,
and the total loss fraction is systematically lower for the thicker atmospheres.
The drag of the atmosphere as the impactor passes through 
can reduce the erosion both immediately and 
by mitigating subsequent oscillations and secondary impacts.
For the faster collisions, as before,
the behaviour remains comparatively simple with more immediate erosion
and the results are less affected by the atmosphere's mass
in terms of timing.

\subsection{Convergence and Other Tests} \label{sec:results:tests}

To study the results of using different particle numbers,
we duplicated each of the two slower fiducial simulations
and the $10^{-2}$~$M_\oplus$-atmosphere fast ones with 
$10^6$, $10^{6.5}$, $10^{7.5}$, and $10^8$ SPH particles (per Earth mass).
For this initial project, we used $10^7$ particles for the main suite
to explore this new parameter space.
\citet{Kegerreis+2019} showed that $10^7$ particles 
are approximately the minimum number required 
to resolve all of the major processes in sufficient detail.
That being said, for the atmospheric-erosion tests specifically
(for thicker atmospheres than here),
lower particle numbers still yielded results 
within 10\% of the converged value.

Fig.~\ref{fig:m_unb_evol_m_atm} shows that 
the number of particles required for convergence clearly depends on the scenario
in addition to the atmosphere mass.
The thicker atmospheres appear well converged 
by only $10^{6.5}$ particles,
as are the $10^{-2}$~$M_\oplus$ atmospheres 
for the high-speed scenarios.
For the thinner atmospheres in the slower scenarios,
the final results differ by a few percent 
even between $10^{7.5}$ and $10^{8}$ particles.
As found by \citet{Kegerreis+2019}, 
the discrepancies manifest primarily after the
initial impact, when debris falls back in 
and other smaller-scale processes can affect the overall results.
Furthermore, regions where the atmosphere is only partially lost 
require many layers of particles to resolve,
which is exacerbated when additional atmosphere 
is eroded multiple times after the initial shock.
While this lack of perfect convergence is important to note,
we can constrain the resulting systematic uncertainty for 
the loss fraction across the suite of $10^{-2}$~$M_\oplus$ atmospheres
to around 2\% in slow scenarios
and much smaller in more violent cases.

In order to further test the dependence of our results 
on the type of finite-particle issues discussed 
in relation to the slow, head-on collision 
in the \nth{1} row of Fig.~\ref{fig:fiducial_snaps},
we ran ten duplicate simulations 
with the target rotated to different orientations.
Most of the resulting loss fractions agree
to within a few percent of the mean of 0.47.
However, three produced $\sim$0.07 more fractional erosion 
and one a remarkable 0.21 less, giving a standard deviation of 0.08.
The qualitative evolution appears much the same in all ten cases,
but the details of the fall-back and sloshing
that follows the initial impact and rebound 
(see Fig.~\ref{fig:m_unb_evol_m_atm})
can differ significantly in magnitude.
This chaotic behaviour also helps to explain 
the incomplete convergence of the slow, head-on collisions discussed above.
In contrast, similar rotated re-simulations of fast, grazing impacts 
produced the same results consistently,
indicating that these issues are restricted to the 
most sensitive slow, head-on cases.
Therefore, we warn that significant care must be taken 
when interpreting the results from one-off simulations of slow, head-on impacts, 
even at high resolution. 

We ran two additional tests with higher surface temperatures of 1000 and 2000~K
on the target in an otherwise unchanged fast, grazing impact.
The resulting loss fractions were 0.04 and 0.07 higher 
than the original result of 0.52, respectively.
While the warmer atmospheres with their greater scale heights 
are indeed lost slightly more easily,
this provides additional confidence that it is only a minor effect.

We also re-ran the same fast, grazing impact 
with a target mantle made of basalt \citep[][Table II]{Benz+Asphaug1999}, 
resulting in a slightly cooler, lower density body than the default granite.
The resulting loss was only 0.02 greater than the standard case's 0.52,
with erosion occurring in the same locations at the same times, 
as was also seen in the temperature tests. 
This suggests that the atmospheric loss is not highly sensitive to 
mild changes in the target's material and precise internal structure.

\begin{figure*}[t]
	\centering
	\includegraphics[width=\textwidth, trim={14mm 195mm 14mm 13mm}, clip]{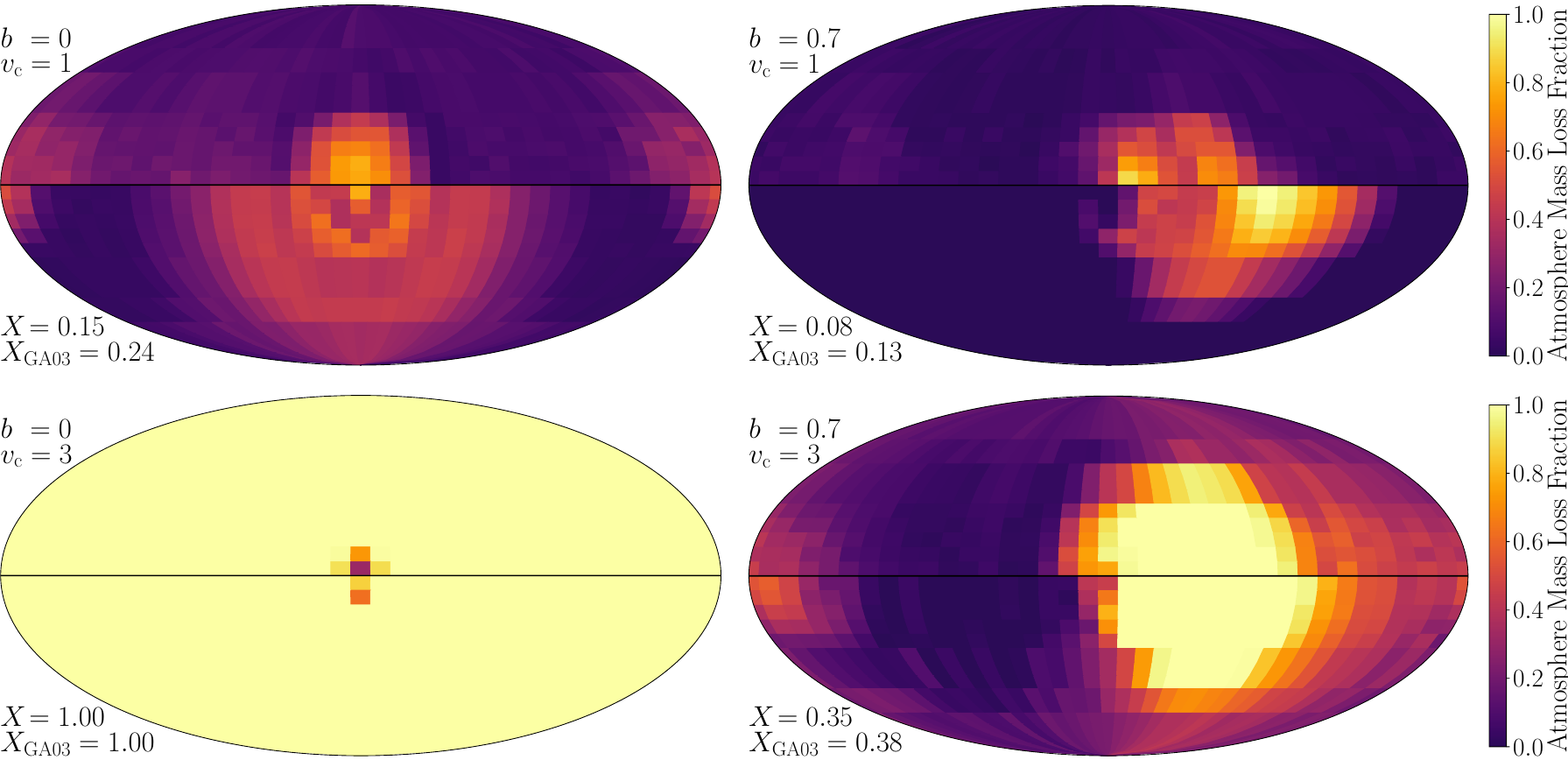}
  \caption{
    The loss fraction of local atmosphere (top hemispheres)
    for the four fiducial simulations,
    on a Mollweide projection as in Fig.~\ref{fig:gnd_vel_peak_time}.
    The bottom hemispheres show the corresponding loss estimates 
    from \citet[][\citetalias{Genda+Abe2003}]{Genda+Abe2003} 
    using the peak ground speeds from our study that are shown in 
    Fig.~\ref{fig:gnd_vel_peak_time}.
    The annotations give the total loss, $X$, from the simulations globally
    for comparison with the total \citetalias{Genda+Abe2003} estimates. 
		\label{fig:X_lat_lon}}
\end{figure*}

\subsection{Ground Speed} \label{sec:results:ground}

The one-dimensional estimates of 
\citet[][hereafter \citetalias{Genda+Abe2003}]{Genda+Abe2003}
predict the local atmospheric loss for a given vertical ground speed.
By defining the `ground' simulation particles as those in the outermost shell 
of the target's mantle,
we can track their movement 
as shock waves (and the impactor itself) perturb them,
as illustrated in Fig.~\ref{fig:gnd_quiver}.
We define longitude~=~$0^\circ$ to be the point of contact
with $\pm 180^\circ$ the antipode,
and latitude~=~$0^\circ$ is the impact ($z=0$) plane.
The maximum outwards radial speed
and the time at which it occurs at each location
are given in Fig.~\ref{fig:gnd_vel_peak_time}
for the four fiducial simulations.

The two head-on impacts are symmetric in longitude
and show high peak speeds near the impact point and the antipode.
For the slower of the two,
the target recoils following the initial collision 
to shoot a plume of material back through the point of impact
and a slightly less dramatic ejection at the antipode,
causing the peak velocities in Fig.~\ref{fig:gnd_vel_peak_time} 
at those longitudes.
Some earlier erosion around the antipode is also 
caused by the initial shock wave,
which is the origin of the maximum velocities 
at most of the other longitudes and latitudes.
As shown in Fig.~\ref{fig:gnd_vel_peak_time},
this occurs a bit less than an hour before the peak recoil.

For the faster head-on collision, 
the impact is more destructive and no such bounce-back plume is seen.
Instead, almost the entire surface is kicked immediately by 
the shock wave to faster than the escape speed, 
explaining the near-total erosion of atmosphere plus some lost mantle 
that was highlighted in Fig.~\ref{fig:fid_will_be_unb}.
In both head-on cases, the lower speeds at high latitudes
simply reflect the rotational symmetry
(as our planets are not spinning).

The two grazing collisions show similar behaviour to each other
with high speeds at positive longitudes, 
i.e. in the path of the impactor as it passes through the point of contact.
The rest of the planet is hit by a shock wave, 
but not one nearly as strong as in the head-on cases,
and with only a mild peak at the antipode.
Unlike the head-on impacts, 
the grazing scenarios are not rotationally symmetric.
Higher latitudes are less affected by the relatively small impactor
and show little longitudinal variation.

In the slower grazing collision,
the local loss around the impact site happens quickly, 
but the peak speeds everywhere else occur up to an hour later,
corresponding to the initial fall-back of some impactor fragments 
and the recoiling oscillation of the planet.
In the faster grazing case, 
the shock wave quickly produces the peak speeds 
across most of the surface,
with little significant fall-back of fragments.
The late times to the positive-longitude side of the impact site 
are less meaningful since most of this material is carried away 
at a roughly constant speed with the surviving impactor,
slightly slower than the impactor's initial 3~$v_{\rm esc}$.
The peak antipode speeds are caused by the violent sloshing of the target
as it begins to resettle following the shock.

Fig.~\ref{fig:gnd_vel_peak} shows a subset of the same peak ground speeds 
for comparison with those predicted by 
\citet[][hereafter \citetalias{Inamdar+Schlichting2015}]{Inamdar+Schlichting2015}.
These are independent of the impact parameter
so nominally correspond to head-on collisions.
They assume that the impactor's momentum is transferred 
at the point of contact and is conserved 
with a constant speed of shocked material within 
the propagating spherical shock wave.
While this inevitably overestimates the ground speed 
close to the point-mass impact,
it also significantly underestimates the peak speed everywhere else
and cannot reproduce the increase in speed at the antipode.
This is unsurprising given their assumption that the entire volume of material
traversed by the shock is all travelling at the same speed.
In reality and in our simulations, 
the shock front moves much faster than the material behind it.
The overprediction near the impact site has little effect on the results 
as all atmosphere is removed there regardless, 
but the low speeds elsewhere 
lead to significant underestimates for the erosion.

\citetalias{Inamdar+Schlichting2015}'s model 
does not include the effects of gravity, the density profile,
rarefaction waves after the shock reaches a surface, 
and the non-zero size and non-instant momentum transfer of the impactor.
The internal structure of the planet 
changes dramatically as the large impactor plunges messily through the mantle;
at high speeds, the impactor can even reach the core of the target 
well before the shock wave has reached the other side.
It is possible that with additional modifications such models 
may be made useful, especially for fast, grazing impacts 
where the shock drives the majority of the loss in a simpler manner, 
though in that case an estimate for the fraction of the impactor's momentum 
that is transferred would also be required,
dependent on the impact angle, speed, and planets' radii.

\begin{figure*}[t]
	\centering
	\includegraphics[width=\textwidth, trim={7mm 8mm 6mm 7mm}, clip]{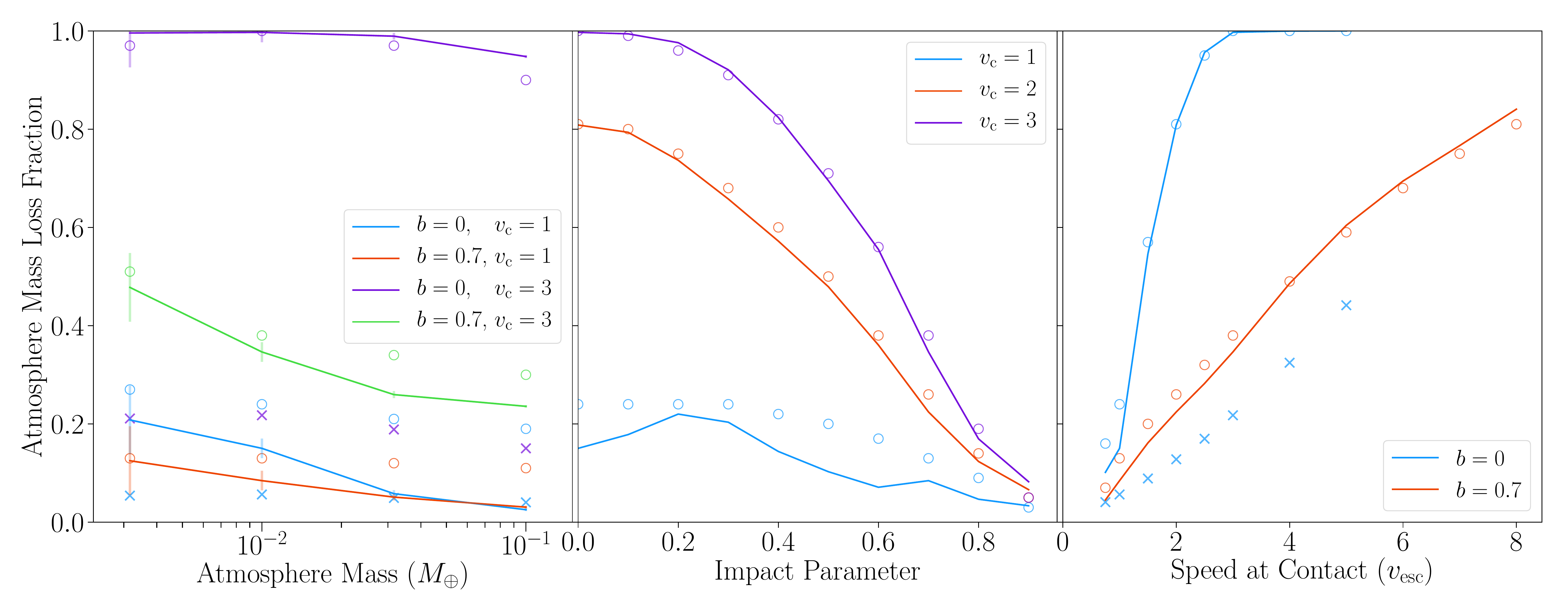}
  \caption{
    The lost mass fraction of the atmosphere for different: 
    (left) atmosphere masses, in each of the fiducial impact scenarios;
    (middle) impact parameters, for three different speeds;
    and (right) speeds at contact, for each fiducial impact parameter;
    all with $\sim$$10^7$ particles.
    The error bars in the left panel 
    show the approximate, conservative uncertainty 
    due to incomplete numerical convergence,
    which becomes significant for the lowest atmosphere mass.
    The circles show the corresponding \citet{Genda+Abe2003}
    estimates based on the peak ground speeds.
    For the head-on collisions,
    the crosses show the \citet{Inamdar+Schlichting2015} estimates
    based solely on the impactor's mass and speed relative to the target
    (their Fig.~4).
		\label{fig:X_sets}}
\end{figure*}

\subsection{Local and Global Atmospheric Loss} \label{sec:results:loss}

Now that we have examined the ground speeds across the planet 
for the fiducial impacts
and introduced 1D and analytical estimates for comparison,
we show in Fig.~\ref{fig:X_lat_lon} the 
local atmospheric mass loss in each region 
for the four fiducial impacts.

The loss fractions broadly follow the distributions of peak ground speeds
in Fig.~\ref{fig:gnd_vel_peak_time},
and the \citetalias{Genda+Abe2003} results based on our peak speeds 
also match well the simulated loss in many places.
Encouragingly, this implies that their 1D calculations 
and our SPH simulations reproduce similar results 
for a ground shock wave eroding the atmosphere above it 
once it arrives at the surface.

This is not always the case
for the more complicated scenarios we are dealing with here.
Perhaps the most significant reason is that for these estimates we have taken 
a single value for the peak ground speed at each location,
whereas in reality the atmosphere can be ejected at many points in time --
as was shown in Fig.~\ref{fig:gnd_vel_peak_time}.
We also cannot fix this simplification by applying \citetalias{Genda+Abe2003}'s
estimates at, for example, all local-in-time maximum ground speeds,
simply because the atmosphere must still be present above 
the ground for a shock to remove it.
After the initial impact, some parts of the atmosphere 
could survive relatively undisturbed and be removed by subsequent shocks.
However, other parts could be partially shocked away 
to fall back down at a later time, 
which may or may not coincide with later shocks.
Thus, the assumption of a single ground speed 
could either over- or underestimate the actual local loss.
We also used the radial ground speeds rather than the total, 
which, if used instead, produce slightly different qualitative results 
but very similar values for the total erosion.

Another important issue is the large size of the impactor
and its complicated interaction with the target,
compared with a simple point-mass explosion 
that would better produce loss just from ground shocks.
Significant amounts of material can thus be ejected directly 
by the impactor ploughing through the atmosphere and mantle,
especially in grazing impacts.

Finally, there are the underlying assumptions 
made and discussed by \citetalias{Genda+Abe2003},
such as their use of an ideal gas EoS
and ignoring lateral motion of the atmosphere,
both of which are likely to be more valid in their targeted regime 
of even thinner atmospheres.
However, the fact that our simulations agree with theirs in many cases 
suggests that these simplifications are often not too important.

The overall results for the suite are presented in Fig.~\ref{fig:X_sets},
showing how the fraction of lost atmosphere varies with atmosphere mass,
impact parameter, and speed.
We find that, unsurprisingly, 
more atmosphere is usually lost from smaller atmospheres, 
more-head-on collisions, and higher speeds.
However, for slower collisions, 
the loss is not a monotonic function of the impact parameter,
and a head-on collision does not cause the most erosion.
By hitting slightly off-centre, 
the impactor can both deliver a strong shock through the planet 
while also encountering and eroding more atmosphere directly.
Although more-grazing impacts can directly remove even more local material,
they fail to deposit enough energy into the shock 
to erode as much atmosphere on the far side.

Apart from this, 
by following the same ground-speed analysis as for the fiducial impacts,
the \citetalias{Genda+Abe2003} estimates 
continue to reproduce the results well in most cases.
As indicated by the ground speeds in Fig.~\ref{fig:gnd_vel_peak}, 
the estimates from \citetalias{Inamdar+Schlichting2015}  
predict far less loss than most head-on collisions.

Bearing in mind that the results for the smallest atmospheres  
are not fully converged numerically,
we find a relatively mild dependence on the initial atmosphere mass,
partly depending on the specific scenario.
This is supported by the good agreement of the 
\citetalias{Genda+Abe2003} estimates, 
which assumed a much thinner atmosphere than ours
along the lines of the Earth's present-day, 
$\sim$$10^{-6}$~$M_\oplus$ atmosphere.

\begin{figure}[t]
	\centering
	\includegraphics[width=0.95\columnwidth, trim={7mm 7mm 7mm 6mm}, clip]{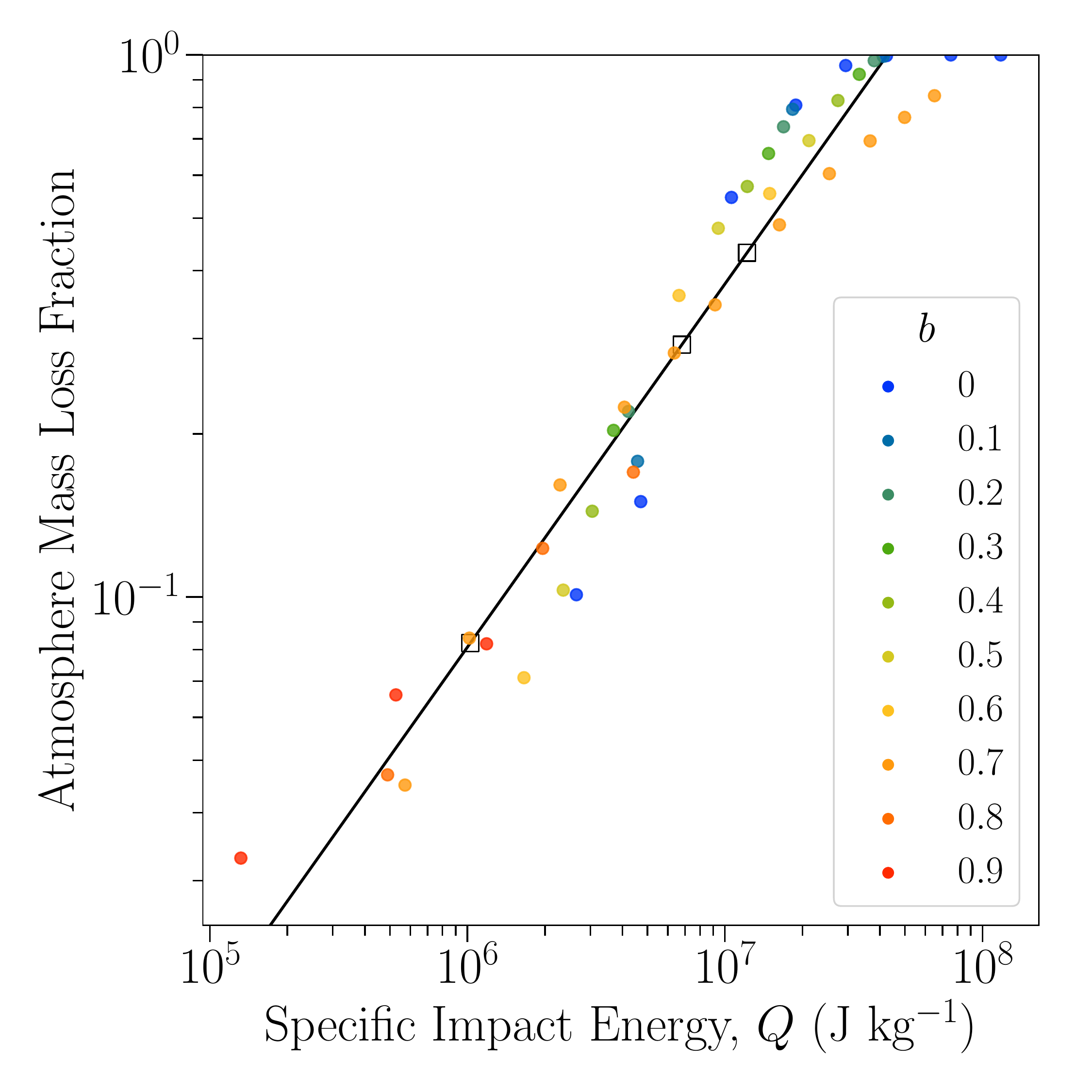}
  \caption{
    The lost mass fraction of the atmosphere for all the simulation scenarios
    as a function of their modified specific impact energy
    (Eqn.~\ref{eqn:Q_mod}),
    coloured by their impact parameter.
    The black line shows our power-law fit (Eqn.~\ref{eqn:X_Q}).
    The lower black square corresponds to the canonical Moon-forming impact
    \citep{Canup+Asphaug2001},
    and the other two to more recent, higher energy scenarios 
    \citep{Cuk+Stewart2012,Lock+2018}.
    These results are also presented numerically in Table~\ref{tab:X_results}.
		\label{fig:X_set_Q}}
\end{figure}

In spite of the complicated details, 
including significant non-monotonic dependence on the angle at low speeds,
we find that a single parameter can be used 
to estimate the erosion from any scenario.
Fig.~\ref{fig:X_set_Q} shows the fraction of atmosphere lost, $X$, 
as a function of the modified specific impact energy,
based on the specific energy used by \cite{Leinhardt+Stewart2012}
to predict disruption.
We find that an additional factor of $(1-b)^2\,(1+2b)$
broadly accounts for the variation across the full range of 
head-on to highly grazing collisions:
\begin{equation}
  Q = (1-b)^2\,(1+2b)\, \tfrac{1}{2}\, \mu_{\rm r}\, v_{\rm c}^2 \, 
    / \, M_{\rm tot} \;,
  \label{eqn:Q_mod}
\end{equation}
where $v_{\rm c}$ is here the SI value not normalised by the escape speed,
and $\mu_{\rm r} \equiv M_{\rm i}M_{\rm t}/M_{\rm tot}$ is the reduced mass.
The added term loosely accounts for the fractional volume 
of the two bodies that interacts:
the volume of the target cap above the lowest point of the impactor at contact,
plus the volume of the impactor cap below the highest point of the target,
divided by the total volume, is
$\tfrac{1}{4}[(R_{\rm t} + R_{\rm i})^3 / (R_{\rm t}^3 + R_{\rm i}^3)] (1-b)^2 (1+2b)$
(see Appx.~\ref{sec:interacting_volume}).
In reality, this is not the exact volume of material that actually interacts,
especially for low-speed collisions or smaller impactors.
Nonetheless, we find empirically that it allows 
a simple power-law fit for the loss fraction to be made in this regime, 
as shown in Fig.~\ref{fig:X_set_Q}:
\begin{equation}
  X \,\approx\, 7.72\!\times\!10^{-6}\, \left(Q \,/\, \rm{J~kg^{-1}}\right)^{0.67} \;, \label{eqn:X_Q}
\end{equation}
capped at one for total erosion.
Note that the effects of changing the impact angle may have an additional 
dependence on the impactor's mass and radius,
so from this initial study alone we can only be certain of 
this scaling law's applicability to bodies of this size.
Its potential extrapolation to wider scenarios 
will be examined in a future study (Kegerreis et al., \emph{in prep}).

\section{Conclusions} \label{sec:conc}

We have presented 3D simulations 
of giant impacts onto terrestrial planets with thin atmospheres.
We explored a wide variety of speeds and impact angles,
as well as a small range of atmosphere masses,
and found a simple scaling law to estimate 
the fraction of atmosphere lost in this regime
of approximately Earth-mass targets and Mars-mass impactors.

Several different processes can dominate the atmospheric loss 
in different scenarios,
depending on, for example, 
whether the impactor can deliver a strong shock wave to remove atmosphere 
on the far side,
or whether impactor fragments fall back after the initial collision.
The interplay of these and other processes 
affects the total fraction of eroded atmosphere,
the local distribution of where atmosphere is lost,
and the time at which it is removed.

For head-on collisions, there is a rapid change with increasing impact speed 
from very little erosion to total loss.
However, for grazing impacts with changing speed --
or for fixed speeds with changing impact angle --
there is a much more gradual transition of partial erosion
that also displays complex,
non-monotonic behaviour at low-to-medium impact parameters.

We find that numerical convergence 
can require many more than $10^6$ SPH particles,
with a strong dependence on the specific impact scenario 
and the measurement in question,
consistently with \citet{Kegerreis+2019}.
The majority of our simulations used $\sim$$10^7$ particles,
which agree with simulations using $10^{7.5}$ and $10^8$
on the fraction of atmosphere lost 
to within a few percent,
with complete convergence in high-speed scenarios where more atmosphere is lost.
Low-speed head-on collisions are particularly chaotic even at high resolution,
making convergence harder to achieve.
We conclude that bespoke convergence tests continue to be crucial 
for any project using planetary SPH simulations.
That being said, our results provide the rough rule of thumb that 
about ten layers of SPH particles are required to model the evolution 
of an atmosphere in these types of scenarios.

By tracking the ground movement throughout the simulations, 
we compared these 3D results with 
\citet[][\citetalias{Inamdar+Schlichting2015}]{Inamdar+Schlichting2015}'s 
analytical estimates for the propagation of shocks from a giant impact,
\citet[][\citetalias{Genda+Abe2003}]{Genda+Abe2003}'s 
1D models for local shock-driven erosion,
and \citetalias{Inamdar+Schlichting2015}'s combined predictions 
for the global loss in a given scenario.
\citetalias{Inamdar+Schlichting2015}'s ground velocities
significantly underestimate the maximum ground speeds in head-on impacts 
owing to the dramatic deformation of the planet
and violent post-impact oscillations.
For the same reasons,
their global predictions underestimate the total loss.
Using our simulated ground speeds, 
\citetalias{Genda+Abe2003}'s estimates match the localised loss fractions well
in most cases, especially when the direct encounter of the impactor 
with the atmosphere is not too important.

In the context of the Earth and the canonical Moon-forming impact, 
only around 10\% of the atmosphere would have been lost
from the immediate effects of the collision.
This suggests that the canonical impact itself cannot single-handedly 
explain the discrepancies between the volatile abundances 
of the Earth and chondrites by eroding the early atmosphere,
compared with alternative, more-violent Moon-forming scenarios. 
However, the caveat of `immediate' erosion is important,
because we have here only considered the direct, dynamical 
consequences of a giant impact.
As examined by \citet{Biersteker+Schlichting2019}, 
the thermal effects of a giant impact heating the planet 
might erode comparable atmosphere to that ejected by shocks,
though the volatile loss may not be that efficient 
even from a hot post-impact disk \citep{Nakajima+Stevenson2018}.
In addition, we took the simple approach here of defining
`lost' atmosphere by particles that become gravitationally unbound,
ignoring the fact that significant material can remain bound
and still be ejected far away from the planet. 
In a real planetary system,
whether by interaction with the solar wind 
or by leaving the target's Hill sphere of gravitational influence,
much of the eroded-but-bound atmosphere could still be lost.
As a separate point, \citet{Genda+Abe2005} showed that 
the presence of an ocean can significantly enhance atmospheric loss,
such that in the canonical Moon-forming scenario,
closer to half the atmosphere could be immediately removed.
Their models combined with our results could be used 
to estimate the amount of an ocean that would be removed
in different scenarios, 
to constrain the extent of fractionation between volatiles.
Future simulation studies could potentially resolve an ocean directly 
and test such erosion in more realistic detail.

The details of atmospheric erosion 
by giant impacts are complicated.
These simulations provide a simple scaling law in this regime
and form a starting point from which to explore 
the vast parameter space in detail.
Promising targets for future study include:
investigations of different impactor and target masses;
extensions to both more massive and even thinner atmospheres;
the inclusion of an atmosphere on the impactor as well as the target;
and testing the dependence on the planets' materials, internal structures, 
and rotation rates.
This way, robust scaling laws could be built up to cover 
the full range of relevant scenarios 
in both our solar system and exoplanet systems
for the loss and delivery of volatiles by giant impacts.

\vspace{-1.6em}
\acknowledgments

We thank the anonymous reviewer for their constructive and insightful comments.
The research in this paper made use of the SWIFT open-source simulation code
\citep[\href{http://www.swiftsim.com}{www.swiftsim.com},][]{Schaller+2018} 
version 0.8.5.
This work was supported by the Science and Technology Facilities Council (STFC)
grant ST/P000541/1, and used the DiRAC Data Centric system at Durham University,
operated by the Institute for Computational Cosmology on behalf of the
STFC DiRAC HPC Facility (www.dirac.ac.uk).
This equipment was funded by
BIS National E-infrastructure capital grant ST/K00042X/1,
STFC capital grants ST/H008519/1 and ST/K00087X/1,
STFC DiRAC Operations grant ST/K003267/1 and Durham University.
DiRAC is part of the National E-Infrastructure.
JAK is supported by the ICC PhD Scholarships Fund
and STFC grants ST/N001494/1 and ST/T002565/1.
RJM is supported by the Royal Society.

%

\vspace{5mm}


\vspace{-1.7em}
\software{
  SWIFT (\href{www.swiftsim.com}{www.swiftsim.com},
  \citet{Kegerreis+2019}, \citet{Schaller+2016}, version 0.8.5);
  SEAGen (\href{https://pypi.org/project/seagen/}{pypi.org/project/seagen/}).
  }



\vspace{16em}
\appendix

\section{Impact Initial Conditions}
\label{sec:init_cond}

For each scenario, we choose the impact parameter, $b=\sin(\beta)$,
and the speed, $v_{\rm c}$, that the impactor would reach 
at first contact with the target,
as illustrated in Fig.~\ref{fig:impact_scenario}. 
The distance between the body centres and the $y$ position at contact are 
\begin{align}
  r_{\rm c} &= R_{\rm i} + R_{\rm t} \\
  y_{\rm c} &= b \, r_{\rm c} \;.
\end{align}
The velocity at infinity,
\begin{equation}
  v_{\rm inf} = \sqrt{v_{\rm c}^2 - 2 \mu / r_c} \;,
\end{equation}
is zero for a parabolic orbit when $v_{\rm c}=v_{\rm esc}$,
where $\mu = G (M_{\rm t} + M_{\rm i})$ is the standard gravitational parameter
and $v_{\rm esc}$ is the two-body escape speed.
Note that for targets with atmospheres, 
we account for the mass of the atmosphere but ignore its thickness.

For elliptical or hyperbolic orbits, 
the speed and $y$ position at any earlier time can be calculated 
using the vis-viva equation and conservation of angular momentum,
where $y$ is in the rotated reference frame where $v$ is in the $x$ direction.
\begin{align}
  a &= \left(\dfrac{2}{r_{\rm c}} - \dfrac{v_{\rm c}^2}{\mu}\right)^{-1} \\
  v &= \sqrt{\mu \left(\dfrac{2}{r} - \dfrac{1}{a}\right)} \\
  y &= \dfrac{y_{\rm c} v_{\rm c}}{v} \label{eqn:cnsv_ang_mom} \;,
\end{align}
where $a$ is the semi-major axis, which is negative for hyperbolic orbits,
and $v_{\rm c}$ is the speed at contact.

In order to rotate the coordinate system such that, at contact, 
the velocity will be in the $x$ direction (a purely aesthetic choice),
we first find the periapsis, $r_{\rm p}$, and then the eccentricity, $e$.
Taking the vis-viva equation at periapsis and 
using Eqn.~\ref{eqn:cnsv_ang_mom} to eliminate the speed gives
\begin{align}
  r_{\rm p}^2 &- a\, r_{\rm p} + 
    \tfrac{a v_{\rm c}^2 y_{\rm c}^2}{2 \mu} = 0 \\
  r_{\rm p} &= \dfrac{a \pm 
    \sqrt{a^2 - \tfrac{2 a v_{\rm c}^2 y_{\rm c}^2}{\mu}}}{2} \\
  e &= 1 - r_{\rm p} / a \;,
\end{align}
which allows calculation of the true anomaly 
(in this case its complement, $\theta$)
and the angle of the velocity away from the radial vector, $\alpha'$:
\begin{align}
  \theta &= \cos^{-1} \left(\dfrac{1 - \tfrac{a\left(1-e^2\right)}{r}}{e}\right) \\
  \alpha' &= \sin^{-1} \left(\dfrac{a^2 \left(1 - e^2\right)}{2ar - r^2}\right) \;.
\end{align}
The final angle needed to rotate the starting $x$, $y$, and $v$ is
\begin{equation}
  \phi = \alpha' - \theta_{\rm c} + \theta - \sin^{-1}\tfrac{y}{r} \;,
\end{equation}
where the c subscript again signifies at contact.
In the special case of a parabolic orbit,
the contact and initial speeds and angles can be calculated directly:
\begin{align}
  v &= \sqrt{2 \mu / r} \\
  \theta &= \pi - \cos^{-1} \left(\dfrac{y_{\rm c}^2 v_{\rm c}^2}{\mu r}\right) \\
  \alpha' &= \theta_{\rm c} / 2 \;,
\end{align}
followed by the same rotation by $\phi$.

The time taken from the initial position to contact, $t_{\rm c}$, 
can be found by using the eccentric anomaly, $E$,
\begin{align}
  E_{\rm ell} &= \cos^{-1} \dfrac{e + \cos \theta}{1 + e \cos \theta} \\
  E_{\rm hyp} &= \cosh^{-1} \dfrac{e + \cos \theta}{1 + e \cos \theta} \\
  E_{\rm par} &= \tan \tfrac{\theta}{2}
\end{align} 
and mean anomaly, $M$, 
\begin{align}
  M_{\rm ell} &= E - e \sin E \\
  M_{\rm hyp} &= -E + e \sinh E \\
  M_{\rm par} &= E + E^3/3
\end{align} 
to find the time since periapsis, $t_{\rm p}$,
\begin{align}
  t_{\rm p, ell, hyp} &= \sqrt{\dfrac{|a|^3}{\mu}} \,M \\
  t_{\rm p, par} &= \sqrt{\dfrac{2 \, r_{\rm p}^3}{\mu}} \,M \;.
\end{align}
Then, $t_{\rm c} = t_{\rm p}(\theta) - t_{\rm p}(\theta_{\rm c})$.

For a radial orbit, the time until the point masses would contact, 
$t_{\rm p}'$, is
\begin{align}
  t_{\rm p, par}' &= \sqrt{\dfrac{2 r^3}{9 \mu}} \\ 
  t_{\rm p, ell}' &= \dfrac{\sin^{-1} \left(\sqrt{w r}\right) - \sqrt{w r (1 - w r)}}
    {\sqrt{2 \mu w^3}} \\
  t_{\rm p, hyp}' &= \left[\sqrt{(|w| r)^2 + |w| r}\right. \\
    &\phantom{=}\;\;
    \dfrac{\;\left. - \ln\left(\sqrt{|w| r} + \sqrt{1 + |w| r}\right)\right]}
    {\sqrt{2 \mu |w|^3}} \;,
\end{align}
where $w$ is the standard constant
\begin{align}
  w &= \dfrac{1}{r} - \dfrac{v^2}{2\mu} \;,
\end{align}
and we can extract $t_{\rm c}$ as before.

For these simulations, we choose the time until contact to be 1~hour
to determine the initial separation and positions.
The equations for the mean anomaly in terms of the eccentric anomaly 
do not have analytical inversions,
so we simply iterate an estimate of the initial 
separation, $r$, until we obtain the desired $t_{\rm c}$.

\section{Approximate Interacting Volume}
\label{sec:interacting_volume}

For this crude estimate of the fractional volume of the target and impactor 
that interacts in a grazing collision, 
we consider the situation illustrated in Fig.~\ref{fig:impact_scenario}
by the target and dotted-line impactor at contact.
We assume that the relevant portion of the target is the spherical cap 
above the horizontal plane set by the lowest point of the impactor at contact.
The relevant portion of the impactor is that below the horizontal plane
set by the highest point of the target.

The volume of a spherical cap with height $d$ is 
\begin{equation}
  V_{\rm cap}(R, d) = \tfrac{\rm \pi}{3} d^2 \left(3 R - d\right) \;,
\end{equation}
where $R$ is the radius of the sphere.
In our scenario, the heights of both caps are 
\begin{equation}
  d = R_{\rm tot} - R_{\rm tot} \sin\beta  = R_{\rm tot} (1 - b) \;,
\end{equation}
where $R_{\rm tot} = R_{\rm t} + R_{\rm i}$ 
is the distance between the two centres.

The interacting volume is then the sum of the target and impactor caps:
\begin{align}
  V &= V_{\rm cap}(R_{\rm t}, d) + V_{\rm cap}(R_{\rm i}, d) \nonumber\\
    &= \tfrac{\rm \pi}{3} d^2 \left[ 
      \left(3 R_{\rm t} - d\right) + \left(3 R_{\rm i} - d\right) \right] 
      \nonumber\\
    &= \tfrac{\rm \pi}{3} d^2 \left[ 3 R_{\rm tot} - 2 d \right] \nonumber\\ 
    &= \tfrac{\rm \pi}{3} R_{\rm tot}^2 \,(1 - b)^2\,
      \left[ 3 R_{\rm tot} - 2 R_{\rm tot} \,(1 - b) \right] \nonumber\\ 
    &= \tfrac{\rm \pi}{3} R_{\rm tot}^3 \,(1 - b)^2 \,(1 + 2 b) \;.
\end{align}
Dividing by the total volume of both spheres gives the fractional volume:
\begin{equation}
  f_V = \dfrac{1}{4} \dfrac{R_{\rm tot}^3}{R_{\rm t}^3 + R_{\rm i}^3}
    \,(1 - b)^2 \,(1 + 2 b) \;.
  \label{eqn:v_interact}
\end{equation}
If the impactor is small and the angle is low, 
then the top of the impactor drops below the top of the target when
$d > 2 R_{\rm i} \rightarrow b < 1 - 2 R_{\rm i} / R_{\rm tot}$.
In this case, the full sphere of the impactor should be included 
and the small target cap above the top of the impactor should be removed.
However, this has negligible effect for large impactors
because the discrepancies in the impactor and target volumes almost cancel out.
It starts to become relevant as the impactor's radius falls to
below half of the target's radius,
but even for ${R_{\rm i} = 0.3\,R_{\rm t}}$ the difference is still only 15\%.
The large impactors in this study have ${R_{\rm i} \approx 0.6\,R_{\rm t}}$.
Future work should determine whether this scaling law 
must be modified for different bodies.

It is debatable whether this volume is a sensible 
estimate for low-angle collisions, where it could be argued that 
the entirety of both bodies is involved.
We also note that this is not the actual intersection of two spheres 
if they were to pass through each other.
Under the assumption that the impactor moves in a straight line, 
which is valid for high-speed collisions,
the relevant volume would be given by the intersection 
of a sphere and a cylinder.
However, this is much more difficult to calculate
and seems unlikely to provide much better results 
in what would still be a highly simplified model of a real collision,
which involves dramatic distortion of the colliding `spheres'.
Therefore, we stick to the convenient expression 
and empirical adequacy of Eqn.~\ref{eqn:v_interact}.

\begin{table}[b]
	\begin{center} \begin{tabular}{ccccccccccc}
    \cline{0-2} \cline{5-7} \cline{9-11}
    $b$ & $v_{\rm c}$ & $X$ & & $b$ & $v_{\rm c}$ & $X$ & & $b$ & $v_{\rm c}$ & $X$ \\
		\cline{0-2} \cline{5-7} \cline{9-11}
    0.0 & 1 & 0.150 &  & 0.0 & 2 & 0.808 &  & 0.0 & 3 & 0.997 \\
    0.1 & 1 & 0.178 &  & 0.1 & 2 & 0.794 &  & 0.1 & 3 & 0.994 \\
    0.2 & 1 & 0.220 &  & 0.2 & 2 & 0.737 &  & 0.2 & 3 & 0.976 \\
    0.3 & 1 & 0.203 &  & 0.3 & 2 & 0.658 &  & 0.3 & 3 & 0.921 \\
    0.4 & 1 & 0.144 &  & 0.4 & 2 & 0.572 &  & 0.4 & 3 & 0.824 \\
    0.5 & 1 & 0.103 &  & 0.5 & 2 & 0.479 &  & 0.5 & 3 & 0.695 \\
    0.6 & 1 & 0.071 &  & 0.6 & 2 & 0.360 &  & 0.6 & 3 & 0.555 \\
    0.7 & 1 & 0.084 &  & 0.7 & 2 & 0.224 &  & 0.7 & 3 & 0.346 \\
    0.8 & 1 & 0.047 &  & 0.8 & 2 & 0.123 &  & 0.8 & 3 & 0.170 \\
    0.9 & 1 & 0.033 &  & 0.9 & 2 & 0.066 &  & 0.9 & 3 & 0.082 \\
		\cdashline{0-2} \cdashline{5-7} \cdashline{9-11}
    0.0 & 0 & 0.101 &  & 0.7 & 0 & 0.045 &  & 0.7 & 6 & 0.694 \\
    0.0 & 1 & 0.546 &  & 0.7 & 1 & 0.161 &  & 0.7 & 7 & 0.767 \\
    0.0 & 2 & 0.956 &  & 0.7 & 2 & 0.282 &  & 0.7 & 8 & 0.841 \\
    0.0 & 4 & 1.000 &  & 0.7 & 4 & 0.486 \\
    0.0 & 5 & 1.000 &  & 0.7 & 5 & 0.604 \\
		\cline{0-2} \cline{5-7} \cline{9-11}
	\end{tabular} \end{center}
	\caption{The impact parameter, 
    speed at contact in units of the mutual escape speed,
    and lost mass fraction of the atmosphere 
    for the suite of simulation scenarios, 
    as presented in Fig.~\ref{fig:X_set_Q}.
    \label{tab:X_results}}
\end{table}


\bibliography{gihr.bib}{}
\bibliographystyle{aasjournal}



\end{document}